\documentclass[a4paper,fleqn,usenatbib]{mnras}

\usepackage{mathptmx}
\usepackage{mathtools}

\usepackage[T1]{fontenc}
\usepackage{ae,aecompl,pdflscape}

\usepackage{graphicx}	
\usepackage{amsmath}	
\usepackage{amssymb}	
\usepackage{subfig}
\usepackage{float}
\usepackage{epstopdf}

\def\pb{Pa$\beta$}
\def\br{Br$\gamma$}
\def\ha{H$\alpha$}
\def\hb{H$\beta$}
\def\feii{[Fe\,{\sc ii}]}
\def\fej{[Fe\,{\sc ii}]$\lambda 1.25\mu m$}
\def\feh{[Fe\,{\sc ii}]$\lambda 1.64\mu m$}

\def\hm{H$_2$}
\def\hml{H$_2$$\lambda 2.12\mu m$}
\def\p1{Paper~I}

\def\kms {$\rm km\,s^{-1}$}

\title[CNSFRs in Mrk\,42]{Circumnuclear star formation in Mrk~42 mapped with Gemini Near-infrared Integral Field Spectrograph}
\author[Hennig et al.]{Moir\'e G. Hennig$^{1}$\thanks{E-mail: moiregh@gmail.com}, Rogemar A. Riffel$^{1}$ , O. L. Dors$^{2}$, Rogerio Riffel$^{3}$,
\newauthor Thaisa Storchi-Bergmann$^{3}$ and Luis Colina$^{4,5}$\\
$^{1}$ Departamento de F\'\i sica, CCNE, Universidade Federal de Santa Maria, 97105-900, Santa Maria, RS, Brazil \\ 
$^{2}$ Universidade do Vale do Para\'iba, Av. Shishima Hifumi 2911, CEP 12244-000, S\~ao Jos\'e dos Campos, SP, Brazil \\
$^{3}$ Departamento de Astronomia, IF, Universidade Federal do Rio Grande do Sul, CP 15051, 91501-970, Porto Alegre, RS, Brazil \\
$^{4}$ Centro de Astrobiolog\'ia (CAB, CSIC-INTA), Carretera de Ajalvir, 28850 Torrej\'on de Ardoz, Madrid, Spain \\
$^{5}$ ASTRO-UAM, Universidad Aut\'onoma de Madrid (UAM), Unidad Asociada CSIC, Madrid, Spain
}

\date{Accepted XXX. Received YYY; in original form ZZZ}

\pubyear{2016}

\begin{document}
\label{firstpage}
\pagerange{\pageref{firstpage}--\pageref{lastpage}}
\maketitle

\begin{abstract}
We present Gemini Near-infrared Integral Field Spectrograph (NIFS) observations of the inner $1.5\times1.5$ kpc$^2$ of the narrow-line Seyfert 1 galaxy Mrk~42 at a spatial resolution of 60\,pc and spectral resolution of 40\,\kms. 
The emission-line flux and equivalent width maps clearly show a ring of circumnuclear star formation regions (CNSFRs) surrounding the nucleus with radius of $\sim$\,500\,pc. The spectra of some of these regions show molecular absorption features which are probably of CN, TiO or VO, indicating the presence of massive evolved stars in the thermally pulsing asymptotic giant branch (TP-AGB) phase. The gas kinematics of the ring is dominated by rotation in the plane of the galaxy, following the large scale disk geometry, while at the nucleus an additional outflowing component is detected blueshifted by 300-500~\kms, relative to the systemic velocity of the galaxy.
 Based on the equivalent width of \br\, we find evidences of gradients in the age of H\,{\sc ii} regions along the ring of Mrk 42, favoring the pearls on a string scenario of star formation. The broad component of \pb\ emission line presents a Full Width at Half Maximum (FWHM) of $\sim$\,1480\,\kms, implying in a mass of $\sim\,2.5\times10^{6}$~M$_{\odot}$ for the central supermassive black hole. Based on emission-line ratios we conclude that besides the active galactic nucleus, Mrk\,42 presents nuclear Starburst activity.

\end{abstract}



\section{Introduction}

\begin{figure*}
    \includegraphics[scale=0.65]{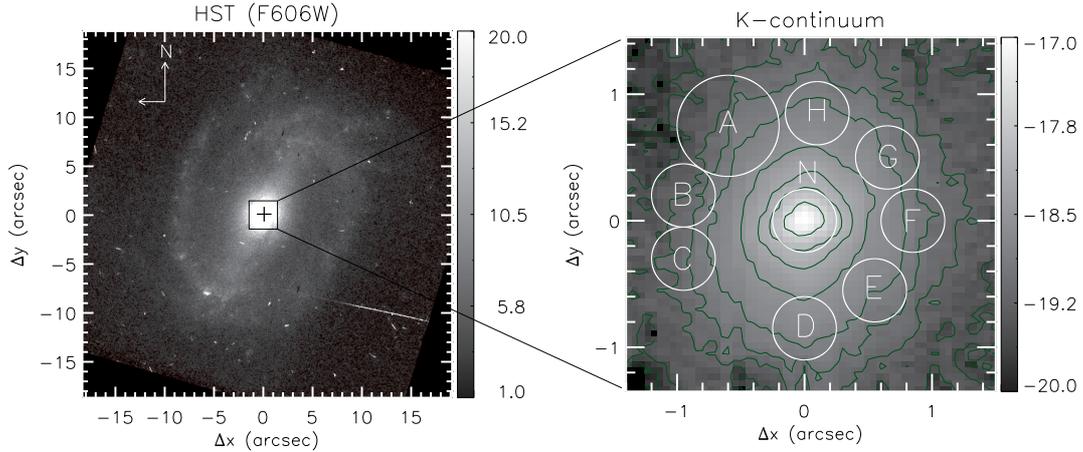}
\caption{The left panel shows an HST image of Mrk 42 obtained through the filter F606W \citep{malkan98}, with the FOV of the NIFS observations indicated by the central square. The right panel shows the K-continuum image obtained from the NIFS datacube. The color bars show the fluxes in arbitrary units for the HST image and in logarithmic units of erg\,\rm cm$^{-2}$\,s$^{-1}$\,\AA$^{-1}$ for the NIFS. The circles correspond to the locations where the CNSFRs are observed, as seen in the Equivalent Width map for the \pb\ emission line in Figure~\ref{flux-ew}. } 
\label{large}
\end{figure*}

Since the early studies by \citet{morgan1958} and \citet{sersic1967} -- see also \citet{kennicutt89} -- that showed that there are many spiral galaxies hosting circumnuclear star formation regions (CNSFRs), much progress has been done in our knowledge about how are these regions originated.
 Inflows of gas from hundred of parsec scales towards the center of galaxies are now commonly observed \citep[e.g.][]{n4051,mrk79,n2110}. These inflows, besides possibly feeding the central supermassive black hole (SMBH) can also trigger (circum-)nuclear star formation, such as it has been observed in some active galaxies.  This relation between nuclear activity and star formation is commonly called as the Active Galactic Nuclei-Starburst  (AGN-Starburst) connection \citep{perry85,norman88}. 
Studies on AGN-Starburst  connections are  of fundamental importance to explain the possible co-evolution of SMBH and its host galaxy, as the star formation in the central region leads to the growth of the bulge mass, while gas accreted by the SMBH leads to the its growth  \citep{Ferrarese2000,Gebhardt2000}.

In barred galaxies the material probably flows through the large scale bars to reach the circum-nuclear region \citep{Sakamoto1999,Jogee2005}, where it can be accumulated and forming a gas reservoir favoring the star formation in circum-nuclear rings with radius of a few hundreds of parsecs \citep{simoes2007,Boker2008}. Inflows of ionized and molecular gas within the inner kiloparsec of nearby active galaxies have also been observed in association to dust spirals using optical \citep[e.g.][]{sm14a,sm14b} and near-infrared \citep[hereafter near-IR, e.g.][]{n4051,mrk79} Integral Field Spectroscopy (IFS). 
Rings of circumnuclear star-formation regions (CNSFRs) are commonly observed in spiral galaxies, being associated to the inner and outer Lindblad resonances  and can be produced by bar or oval distortion perturbations \citep[e.g.][]{schwarz81}. Such rings are more frequently found in AGN hosts in comparison to normal galaxies or starbursts \citep{hunt1999,hunt99b}. In addition, Low-Ionization Nuclear Emission-line Regions (LINERs) hosts show higher fractions of inner rings, while for Seyfert galaxies outer rings are more often observed, with ring fractions being several times larger than in normal galaxies \citep{hunt1999}. Regarding the presence of bars, \citet{hunt1999} found that  active and non-active galaxies show similar incidence of them, thus the distinct fraction of rings seems to be the only morphological difference between active and non-active galaxies.

Several recent studies were aimed in constraining the physical properties of CNSFRs in active and non-active galaxies \citep{davies07,dors08,Boker2008,falcon2014,van-der-lan2015,Riffel2016}. 
 CNSFRs show star-formation rates of 10$^{-3}$--10$^{-1}$ M$_\odot$ yr$^{-1}$ and have ionized gas masses ranging from 10$^4$ to 10$^5$ M$_\odot$ \citep[e.g.][]{dors08,Riffel2016}. 
The scenario that leads to the formation of CNFRs is still an open issue and the way the material settles in the circum-nuclear region may represent different processes for the star formation.  For example, \citet{Knapen2005} considers a distinction between CNSFRs and nuclear star formation regions: in the former, the material that arrives in the central region forms a circum-nuclear ring with $\sim$\,2 kpc radius, while in the latter the material settles in more internal regions (within a few hundred parsecs).
 Two main scenarios have been proposed for the origin of CNSFRs: the \textit{popcorn} and \textit{pearls on a string} scenarios \citep{Boker2008,Elmegreen1994}. In the first scenario, after an increase in the gas density due to accumulation of material in the ring, the stellar clusters are formed at random positions. In this scenario, there are no age sequence for the star clusters along the ring. In the second scenario, the gas flows to the central region along bars and is accumulated at two locations (at the tips of the bar), originating over-dense regions - ODR, in which the stars are formed. Due to the differential rotation of the bar and the disk of the galaxy, the star formation ``moves" along the ring, leading to a sequence of ages for young stellar clusters. 

Until now, studies on star formation in CNSFRs seem to have not produced yet conclusive results about the star formation scenario. In fact, \citet{mazzuca2008}, who used photometric \ha\ data of 22 CNSFRs, found that about half of the rings contain azimuthal age gradients, while \citet{dors08} and \citet{Riffel2016}  did not found any sequence of age in CNSFRs located in three galaxies, using optical and near-IR IFS.

With the goal to improve the knowledge of star formation in CNSFRs, increasing the sample of objects, we have used near-IR IFS to study the circumnuclear star formation and nuclear activity in the nearby active galaxy Mrk~42. This object is classified as an SBa galaxy harboring a Narrow-Line Seyfert 1 nucleus \citep{malkan98}, being located at a redshift $z=0.024634$ \citep{falco1999}, 
corresponding to a distance of 102.6 Mpc by assuming a Hubble constant $H_0= \rm 72\,\rm km\: s^{-1}\: Mpc^{-1}$. At this distance, 1$^{\prime\prime}$ corresponds to 497 pc at the galaxy.
 It harbors a SMBH of mass (0.7-18)$\times$10$^{6}$ as derived from scaling relations and  based on the X-ray excess variance \citep{wang01,bian03,nikolajuk09}. 

\citet{Munoz2007} present an atlas of the central regions of 75 Seyfert galaxies imaged in the near-UV with the Hubble Space telescope (HST), including Mrk~42 in their sample. They found that Mrk 42 has a very bright compact nucleus with a ring of star formation $\approx$\,300 pc of radius, with many stellar clusters individually resolved. Indeed, the observed spiral arm seen in Mrk~42 is very coiled, being described by  \citet{sb08} as  clear ring of star-formation surrounding the nucleus. The size of the ring of CNSFRs observed in Mrk~42 is similar to that derived for the rings in other galaxies, such as NGC\,1068 and NGC\,7469, resulting from a gas ring formed through bar-forcing at  the inner Lindblad resonance \citep[e.g.][]{Wilson}.   
 A broad band HST image of Mrk~42 obtained with the Wide Field Planetary Camera 2 (WFPC2) through the filter F606W shows grand-design nuclear spiral arms, originating from a bar with an extent of 15$^{\prime\prime}$ oriented along position angle $\rm PA=-20/160^\circ$, and  the CNSFRs ring is clearly observed \citep{malkan98,deo06}. The nuclear region of Mrk~42 presents strong 6.2~$\mu$m PAH emission as revealed by Spitzer mid-infrared spectroscopy, with a luminosity of $2\times10^{41}$ erg\,s$^{-1}$, providing strong evidence for intense ongoing star formation in the circumnuclear region \citep{sani10}.

We use J, H and K-band IFS of the inner $3^{\prime\prime}\times3^{\prime\prime}$ ($1.5\times1.5$ kpc$^2$) of Mrk~42 to map, for the first time, its near-IR emission-line flux distributions and kinematics. This paper is organized as follows. In section \ref{obs}, we describe the observations and data reduction procedures. In section \ref{results} we present maps for the fluxes, flux ratios, line-of-sight velocities and velocity dispersions of the molecular and ionized gas emission lines. These maps are discussed in section \ref{discussion} and our conclusions are presented in section \ref{conclusions}.


\begin{figure*}
\begin{center}
 \includegraphics[height=8.4cm]{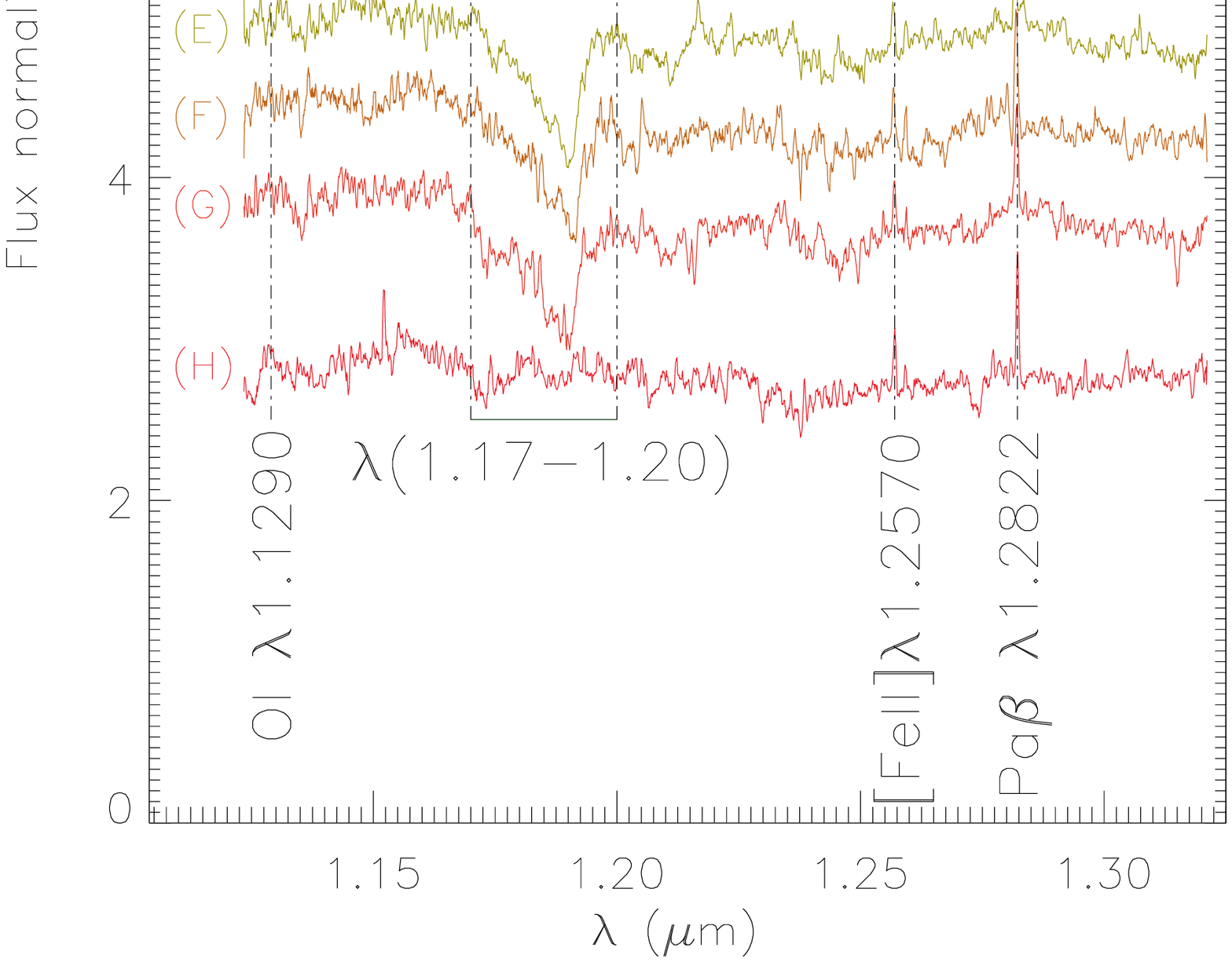}\quad
  \includegraphics[height=8.4cm]{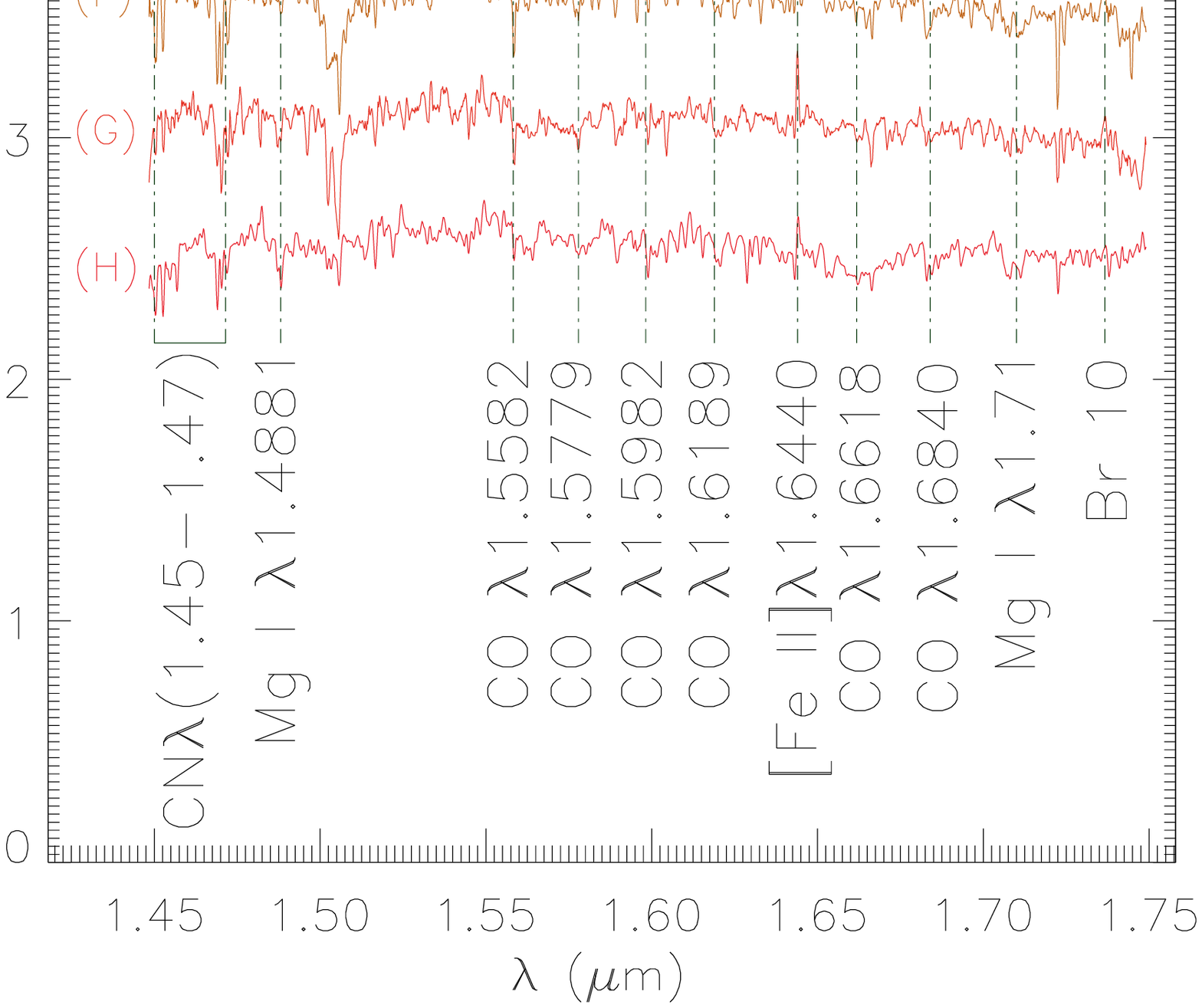} \quad
   \includegraphics[height=8.4cm]{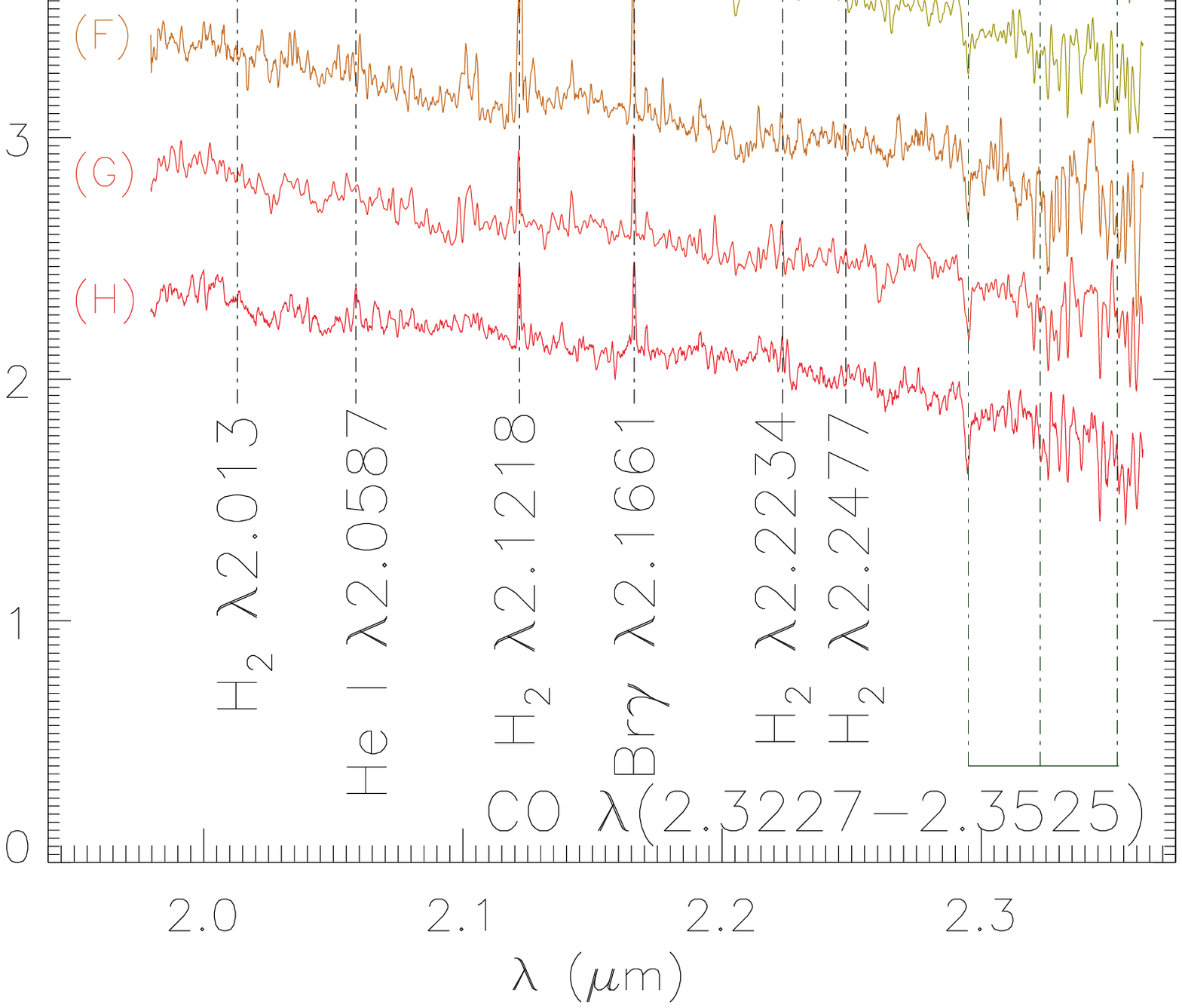} \quad
 \caption{J (left), H (center) and K-band (rigth) spectra for the nucleus (top spectra) and CNSFRs. The spectra were extracted for an aperture of 0$\farcs$25 radius centred at the locations identified by the circles in the bottom-left panel of Fig.~\ref{flux-ew} and are corrected for the Doppler shift. The exception is the location A, for which an aperture of 0$\farcs$4 radius was used. The strongest emission and absorption lines are identified for each band.}
\label{spectra}
\end{center}
\end{figure*}

\section{Observations and data reduction}\label{obs}

The observations of Mrk\,42 were done using the Gemini Near-infrared Integral Field Spectrograph \citep[NIFS,][]{mcgregor03} operating with the ALTAIR adaptive optics module on the Gemini North telescope in May 2014 under the programme GN-2014A-Q-28. The J, H and K-band observations were obtained following the standard Object-Sky-Object dithering sequence, with individual exposures of 520 sec each. We obtained eight on-source exposures for each band, and thus the total integration time is $\sim$\,1.2 hours per band. The NIFS has a square field of view of 3$^{\prime\prime}\times 3^{\prime\prime}$, which was centred at the nucleus of the galaxy during the observations.

For the J-band, we used the filter J\_G5603, resulting in a spectral range from 1.15$\mu$m to 1.35$\mu$m and a spectral resolution of $\approx$\,1.7\AA, as obtained from the Full Width at Half Maximum (FWHM) of typical Ar calibration lamp lines. The H-band observations were centred at 1.65$\mu$m, covering a spectral range between 1.48$\mu$m and 1.80$\mu$m at a spectral resolution of $\approx$\,2.5\AA\ using H\_G5604 filter. 
In K-band the resulted spectral range is 2.01--2.42$\mu$m, centred at 2.20$\mu$m  and the observations were performed using the K\_G5605 filter. The resulting spectral resolution for the K-band is $\approx$\,3.2\AA, as obtained from FWHM of typical ArXe wavelength calibration lamp emission lines.

The data reduction followed the standard procedure of near-IR spectroscopic data treatment and was accomplished using tasks contained in the {\sc gemini.nifs} IRAF\footnote{IRAF is distributed by National Optical Astronomy Observatories, which are operated by the Association of Universities for Research in Astronomy, Inc., under cooperative agreement with the National Science Foundation} package as well as generic IRAF tasks.  
The data reduction procedure included trimming of the images, flat-field correction, sky subtraction, wavelength calibration and s-distortion correction. The telluric absorptions were removed by dividing the spectra of the galaxy by the normalized spectrum of a telluric standard star of A spectral type, observed just before or after the observations of the galaxy. These standards have also been used to flux calibrate the spectra by interpolating a blackbody function to fit the spectrum of each standard star.
 
Finally, individual datacubes for each exposure were created with a spaxel size of 0\farcs05$\times$0\farcs05, which were then combined to a single datacube for each band using the \textit{gemcombine} task of the GEMINI IRAF package. The final datacube for each band contains about 3600 spectra and covers the inner 3$^{\prime\prime}\times 3^{\prime\prime}$ of the galaxy. The angular resolution is about 0\farcs12 for all bands, as obtained from the FWHM of the flux distribution of the telluric standard stars, corresponding to $\approx
$60\,pc at the galaxy. Similar FWHM values are obtained from the flux distributions of the \pb, \br\ and O\,{\sc i}$\lambda1.129\,\mu$m  broad components.

In order to eliminate high frequency noise from the data, but with no loss in the image quality and preserving the flux, we used the IDL script \textit{$bandpass_{-}filter.pro$} to apply a filter of order 3 and a cutoff frequency of 0.35 Ny on all bands. The filter was chosen by comparing the filtered and original continuum image, making sure that there was no removal of intrinsic emission from the galaxy.
 Similar procedures were already applied to NGC\,4303 SINFONI datacubes by our group \citep{Riffel2016} and Butterworth filtering has extensively been used in the treatment of optical IFS of nearby galaxies \citep[e.g.][]{ricci14,may16}.

\section{Results}\label{results}

In the left panel of Figure~\ref{large}, we present an optical image of the inner 40$^{\prime\prime}\times$40$^{\prime\prime}$ of Mrk~42 obtained with the HST WFPC2 through the filter F606W \citep{malkan98}. The central cross marks the position of the nucleus indicated by `+' and the central square indicates the field of view (FOV) of the NIFS observations of 3$^{\prime\prime}\times$3$^{\prime\prime}$. The nuclear grand-design spiral arms and bar are clearly seen in this image, as discussed in \citet{deo06}. The right panel of this figure shows the continuum image of the nuclear region obtained from the NIFS data cube by averaging the fluxes between 2.26 and 2.28 $\mu$m, a region with no strong emission/absorption lines. Although the continuum is more elongated in the direction of the bar, the presence of the bar in the inner 3$^{\prime\prime}\times$3$^{\prime\prime}$ region is not evident. In addition, the K-band continuum image does not show any enhancement at the locations where the CNSFRs were detected in UV continuum images \citep{Munoz2007}, which is an expected result, since young SP are enhanced in the UV spectral region while in the near-IR just a very small fraction of their light is detected \citep{rogerio2011}.
The locations of the CNSFRs are identified from the \pb\ Equivalent Width map shown in the bottom-left panel of Fig.\ref{flux-ew}.

In Fig.~\ref{spectra} we show the J (left panel), H (center panel) and K-band (right panel) spectra for the nucleus and for the CNSFRs. These spectra were extracted within a circular aperture of 0\farcs25 radius, centred at the nucleus and location of each CNSFR, identified as the locations of peak in the Equivalent Width (EqW) map for the \pb\ emission line, shown in the bottom-left panel of Fig.~\ref{flux-ew}.  The nucleus is labeled as `N' and the CNSFRs from `A' to `H' in Fig.~\ref{flux-ew}. The size of the aperture was chosen to include most of the emission form each CNSFR and for region `A', a larger aperture with 0\farcs4 radius was used. The spectra show several emission lines, as \fej, \feh, \pb, \br\ and \hml, as well as absorption features, as the CO band heads at the K and H bands. These spectral features can be used to investigate the origin and kinematics of the emitting gas, as well as physical properties of the stellar populations.

\begin{figure}
\begin{tabular}{cc}
 \includegraphics[scale=0.195]{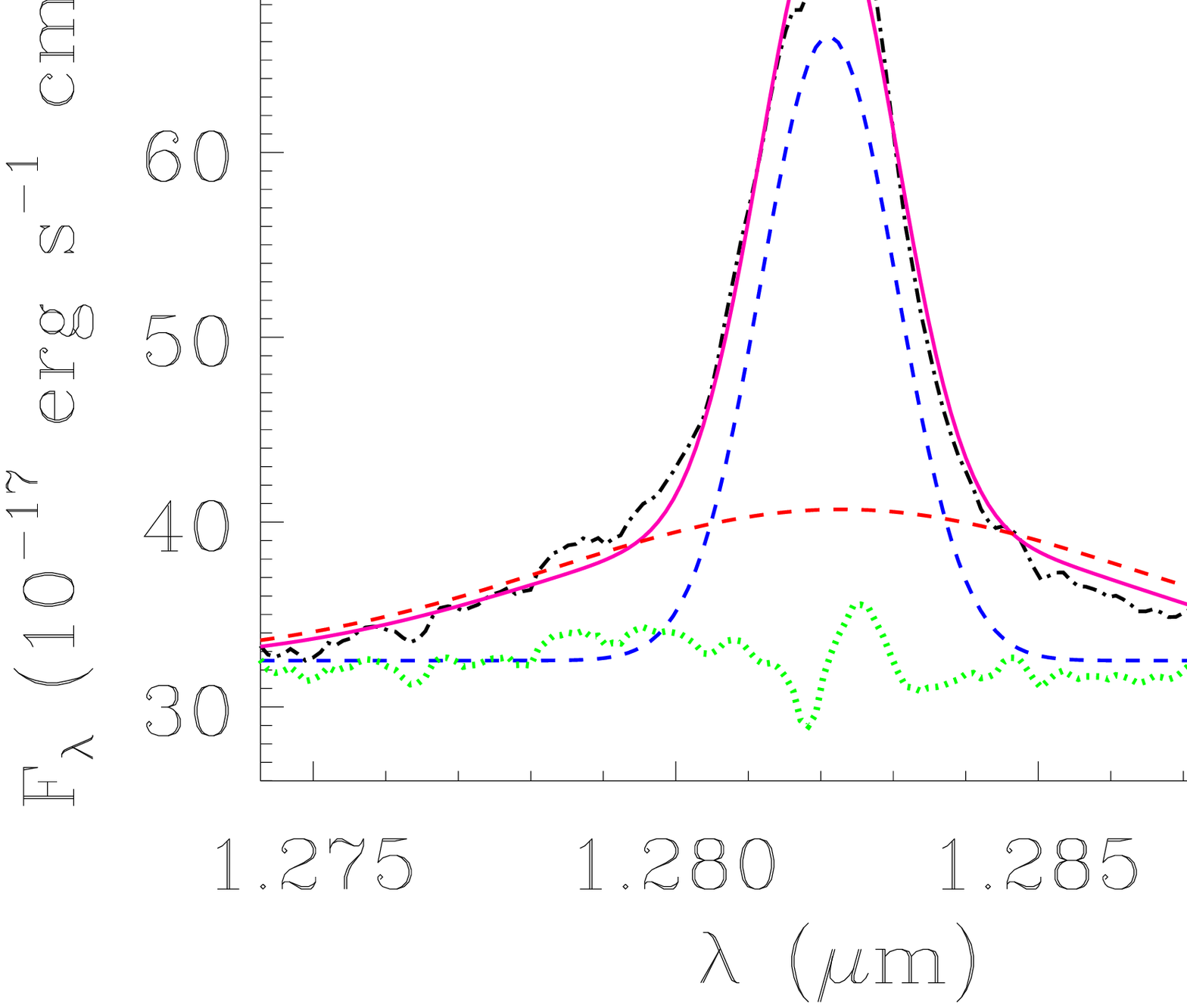}&
    \includegraphics[scale=0.195]{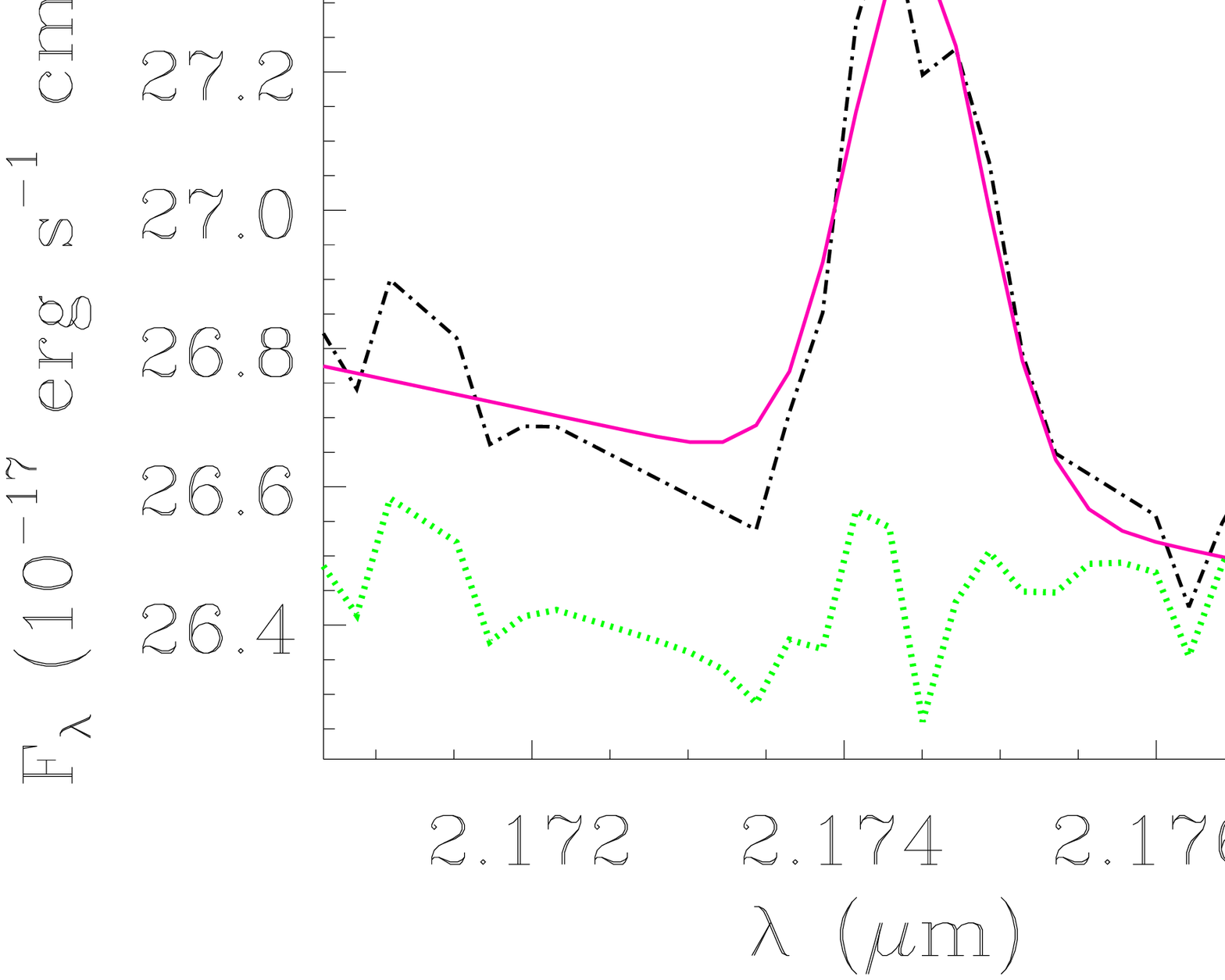} \\
    \includegraphics[scale=0.195]{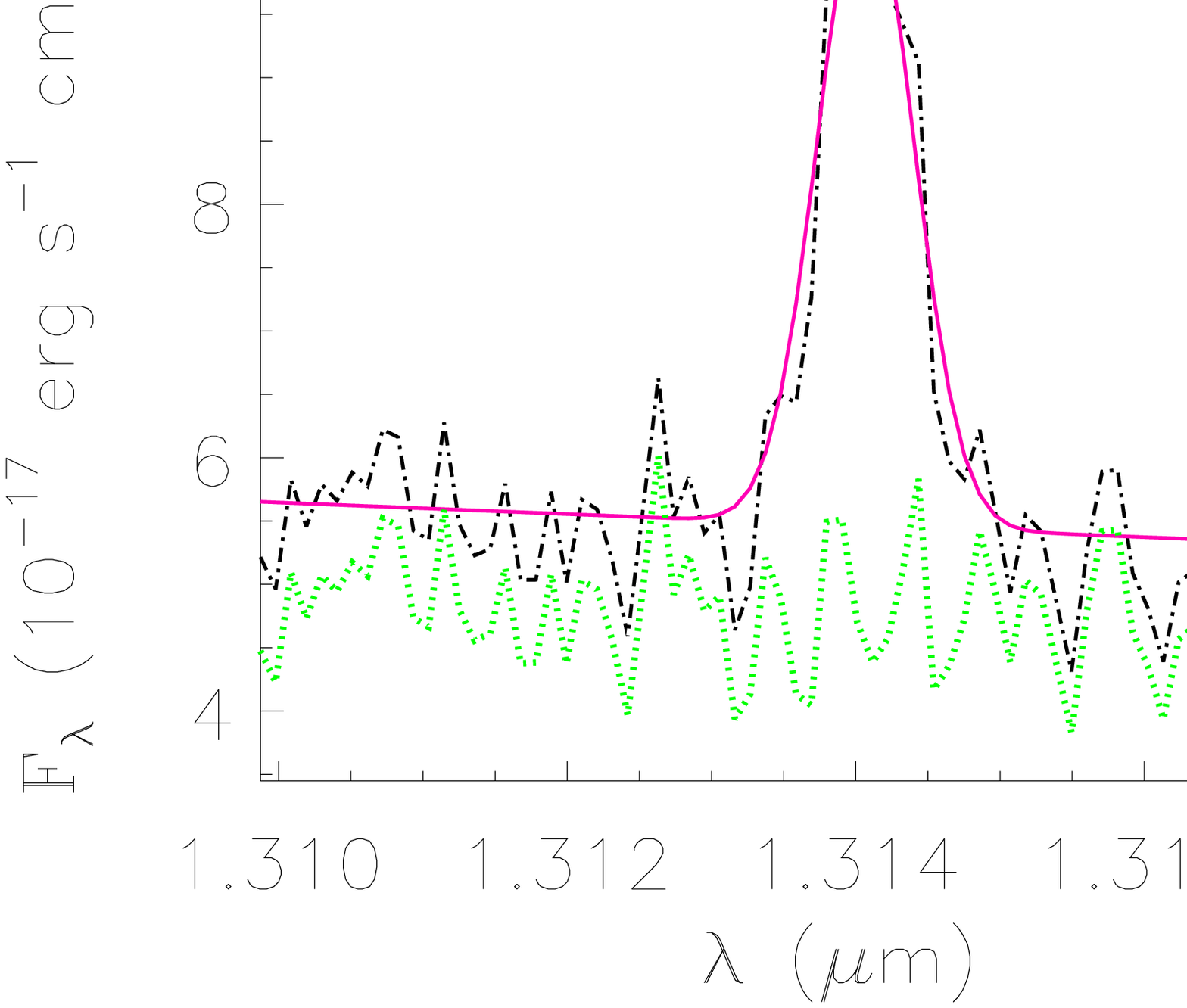}&
     \includegraphics[scale=0.195]{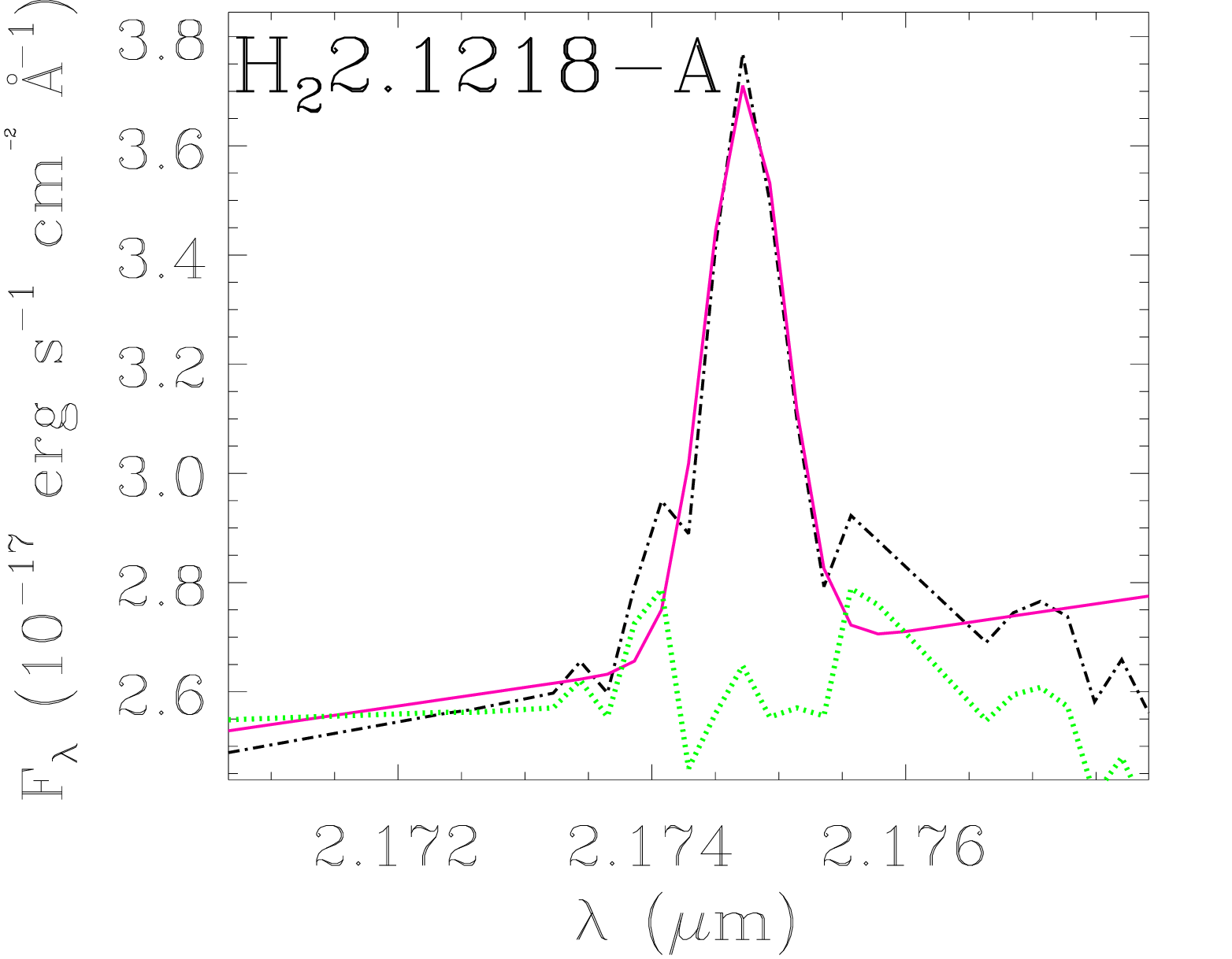} \\

\end{tabular}
\caption{Examples of fits of the emission-line profiles of \pb\ (left) and H$_2$\,$\lambda$\,2.12$\mu$m (right) for the nucleus (top) and region A (bottom), within the apertures identified at the rigth panel of Fig.~\ref{large}. The observed profiles are shown in black, the best fit in magenta and the residuals (plus an arbitrary constant) in green. The \pb\ profile at the nucleus was fitted by two Gaussian components: the broad component is shown as a red dashed line the narrow component as a blue dashed line.}
\label{sample-fit}
\end{figure}

We fitted the profiles of the strongest emission lines by Gaussian curves in order to map their flux distributions and the gas kinematics. The fitting was done using the PROFIT (line-PROfile FITting) routine \citep{profit}.  At most locations the emission line profiles of \hml\, \fej, \feh, \pb\ and \br\ are well reproduced by a single Gaussian function, but for the inner 0\farcs45 radius the recombination lines were fitted with a narrow and a broad component, corresponding to an origin in the Narrow-Line Region (NLR) and Broad-Line Region (BLR), respectively. As the BLR is unresolved, the central wavelength and the width of the broad component were kept fixed at the values obtained by fitting the nuclear profiles shown in Fig.~\ref{spectra}, while the amplitude of the Gaussian was allowed to vary. The line-of-sight velocity ($V_{\rm LOS}$) of the broad component is blueshifted by 190 \kms\ relative to the systemic velocity of the galaxy (as derived in section \ref{disc-kin}) and its FWHM is 1480 \kms, which is larger than the values reported for \hb\ of 670--865\,\kms\ by \citet{bian03} and \citet{wang01}. Examples of fits for the \pb\ and \hml\ profiles are shown in Fig.~\ref{sample-fit} for the nuclear position (top) and for the position of the CNSFR labeled as `A' in Fig.~\ref{flux-ew}. We used the resulting measurements for the emission-line fluxes, $V_{\rm LOS}$ and velocity dispersion ($\sigma$) to construct the two-dimensional maps that are presented in the following sections.

\subsection{Emission-line flux distributions and Equivalent Width maps}\label{FeEqW}

\begin{table*}
\caption{Measured emission line fluxes in units of $10^{-18}$\,erg\,\rm cm$^{-2}$\,s$^{-1}$ and the  E(B-V) values for the positions labeled in Fig.~\ref{flux-ew}, within a circular aperture with radius showed in the second column.}
\begin{tabular}{c c c c c c c c} 
\hline
 &  &  \multicolumn{5}{c}{Fluxes}   &  \\ \hline
 Region & R   & [Fe\,{\sc ii}]$\lambda$\,1.25 & Pa$\beta$ & [Fe\,{\sc ii}]$\lambda$\,1.64  & H$\rm_2$\,$\lambda$\,2.12 & Br$\rm \gamma$ & E(B-V) [\pb/\br] \\ 
\hline
N &  0$\farcs$25  & 105.7$\pm$59.2     & 6\,420.0$\pm$190.0   & 51.4$\pm$25.1      & 112.7$\pm$20.0    & 1\,130.0$\pm$85.0   & 0.06$\pm$0.21   \\ 
A & 0$\farcs$40   & 157.2$\pm$50.1      &  860.0$\pm$36.8      & 176.6$\pm$43.5    & 73.8$\pm$15.8     & 191.4$\pm$29.3      & 0.54$\pm$0.36   \\ 
B & 0$\farcs$25    & 123.1$\pm$10.1      &  258.7$\pm$21.0      & 50.1$\pm$12.7      & 40.5$\pm$9.7     &  66.0$\pm$10.7      & 0.82$\pm$0.41  \\ 
C & 0$\farcs$25   &  46.8$\pm$18.5      &  259.3$\pm$12.9      & 75.0$\pm$18.4      & 40.5$\pm$9.1     &  65.7$\pm$10.4      & 0.81$\pm$0.37   \\ 
D & 0$\farcs$25    &  59.3$\pm$22.5      &  266.2$\pm$14.6      & 51.6$\pm$15.3      & 39.5$\pm$8.1     &  67.4$\pm$12.7      & 0.81$\pm$0.45  \\
E & 0$\farcs$25    &  84.5$\pm$26.4      &  308.7$\pm$15.2      & 63.8$\pm$16.1      & 30.8$\pm$6.3     &  73.3$\pm$8.3       & 0.68$\pm$0.27  \\
F & 0$\farcs$25    &  54.8$\pm$13.4      &  268.0$\pm$16.7      & 77.7$\pm$14.0      & 66.1$\pm$13.5    &  69.3$\pm$8.4       & 0.85$\pm$0.30  \\ 
G & 0$\farcs$25    &  60.1$\pm$23.9      &  282.3$\pm$18.1      & 83.7$\pm$18.9      & 44.2$\pm$7.3     &  61.3$\pm$7.9       & 0.49$\pm$0.32  \\ 
H & 0$\farcs$25    &  79.1$\pm$33.8      &  193.9$\pm$13.8      & 50.8$\pm$14.4      & 50.8$\pm$7.1     &  56.9$\pm$11.3      & 1.11$\pm$0.49  \\ 
 \hline 
\end{tabular}
\label{tab-flux}
\end{table*}

\begin{figure*}
    \includegraphics[scale=0.9]{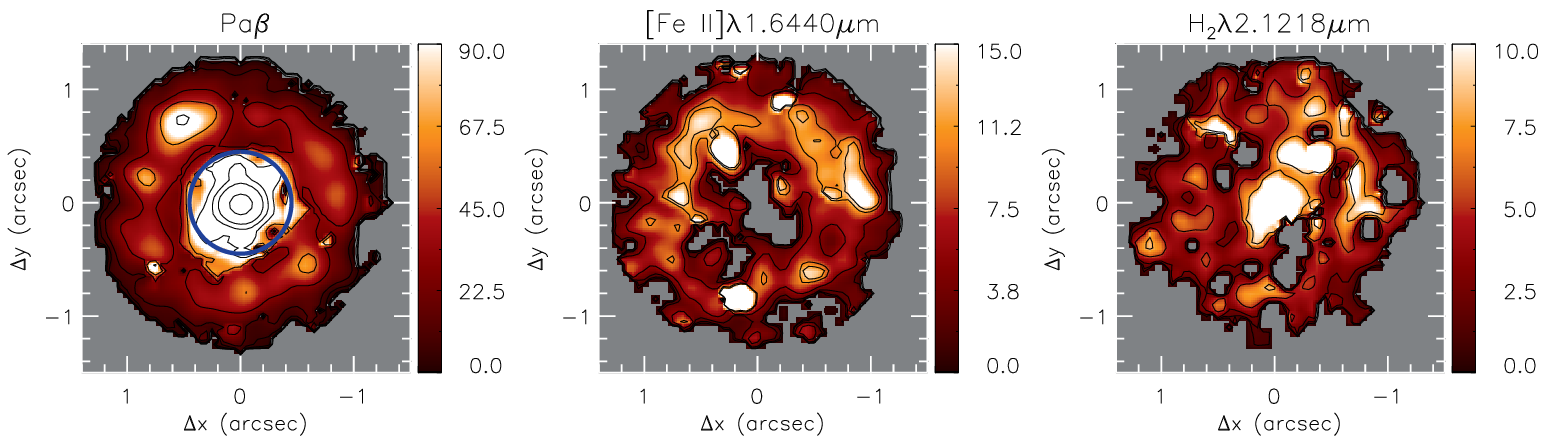}
    \includegraphics[scale=0.9]{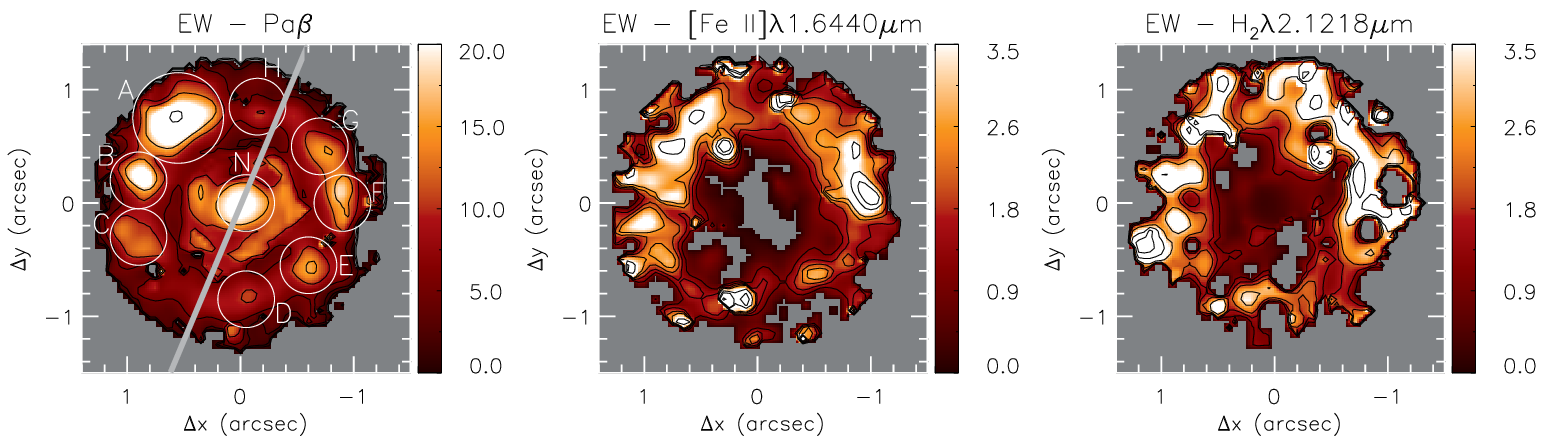}
\caption{Top: Emission-line flux distributions for the \pb\ (left), \feh\ and \hml. The color bars show the fluxes in units of $10^{-19}$\,erg\,\rm cm$^{-2}$\,s$^{-1}$. For Pa$\beta$ emission line, the blue circle around of the nucleus delineates the region (0$\farcs$45 radius) where the \pb\ profile was fitted by two Gaussian curves. 
 Bottom: Equivalent width maps for Pa$\beta$, [Fe\,{\sc ii}]$\,\lambda$\,1.25$\mu$m and H$\rm_2$\,$\lambda$\,2.12$\mu$m emission-lines. The color bars show the EqW in \AA. The circles shown in the Pa$\beta$ EW map correspond to the locations of the CNSFRs and the gray line marks the orientation of the bar, as seen from the HST image of Fig.~\ref{large}.}
\label{flux-ew}
\end{figure*}

In the top panels of Fig.~\ref{flux-ew} we present maps for the \pb\ (left), \feh\ (center) and \hml\ (right) emission line flux distributions. We do not show maps for \fej\ and \br\ as they are very similar to those of \feh\ and \pb, respectively, but noisier. The gray regions in the flux maps represent masked locations where the uncertainties in flux measurements are larger than 40\% or due to the non detection of the emission line.  The blue circle overlaid to the \pb\ flux map delineates the region where two-gaussian components were fitted to the \pb\ line profile at the nuclear region. All maps clearly show the presence of the ring of CNSFRs with radius of 0\farcs6 ($\sim$\,300\,pc), co-spatial with the ring seen in the UV HST image \citep{Munoz2007}, but some differences are seen among the maps. While strongest extra-nuclear \pb\ emission is observed for region `A' at the north-east of the nucleus, the \hm\ and \feii\ emission show stronger emission from regions located at the north-west of the nucleus. In addition, \pb\ show strong nuclear emission, while a weaker nuclear emission is observed in \hm\ and \feii\ is not detected at the nucleus of Mrk\,42.  In Table~\ref{tab-flux} we present flux measurements for the \fej, \pb, \feh, \hml\ and \br\ emission lines for the nucleus and the CNSFRs integrated within a circular aperture of 0\farcs25 radius, except for region `A', for which an aperture of 0\farcs4 was used.

The bottom panels of Fig.~\ref{flux-ew} show the Equivalent Width (EqW) maps for the \pb\ (left), \feh\ (center) and \hml\ (rigth), which can be used to better identify the CNSFRs, as the EqW measures the flux relative to the continuum emission and thus is 'a pure' line emission diagnostic. The white circles overlaid to the \pb\ map (that clearly delineates the CNSFRs ring) marks the locations of the CNSFRs labeled from `A' to `H', as well as the position of the nucleus (`N'). The size of the aperture was chosen to be larger than our spatial resolution bin and includes most of the line emission from the CNSFRs. The gray continuous line represents the orientation of the bar seen in the optical HST image, shown in Fig.~\ref{large}.  The \feh\ and \hml\ show EqW values ranging from nearly zero to $\sim$\,5, while \pb\ shows higher values of up to 14. The highest values of \pb\ EqW  are seen for regions `A' and `B' to the north-east of the nucleus, while the other lines show the highest values both to the north-east and north-west of the nucleus. In addition, the smallest values of EqW for the \hm\ are seen at the nucleus, which may be due to an increase of the continuum emission due to hot dust, as suggested by the nuclear spectrum (see Fig.~\ref{spectra}), which is redder than the extra-nuclear spectra. 

 It should be noticed that the EqW values observed for all emission lines are small and a possible explanation for this is that the continuum contains an important contribution from bulge stars, that dominates the near-IR continuum emission at the central region of galaxies. In order to evaluate the contribution of the stellar bulge continuum, we followed the methodology of \citet{dors08} and performed aperture photometry on the continuum images  to measure the continuum emission at the locations of the CNSFRs and in the surrounding regions. We concluded that the bulge contributes with about 80, 85 and 90\,\% of the total flux in the K, H and J bands respectively. This translates to an increase in EqW values -- after subtracting the contribution from the continuum -- by a factor of 4 for \br\ and \hml\ lines, a factor 6.7 for the \feh\ and a factor of 10 for the \pb\ and \fej\ lines. The EqW values of \pb\ and \br\ emission lines can be even larger if stars formed in previous episodes of star
formation in the ring are present, but evaluating the contribution of these stars requires  high signal-to-noise spectra to allow a proper spectral synthesis and this is beyond the scope of this paper.

\subsection{Emission-line ratios}\label{ratios}

\begin{figure*}
\begin{center}
    \includegraphics[height=4.8cm]{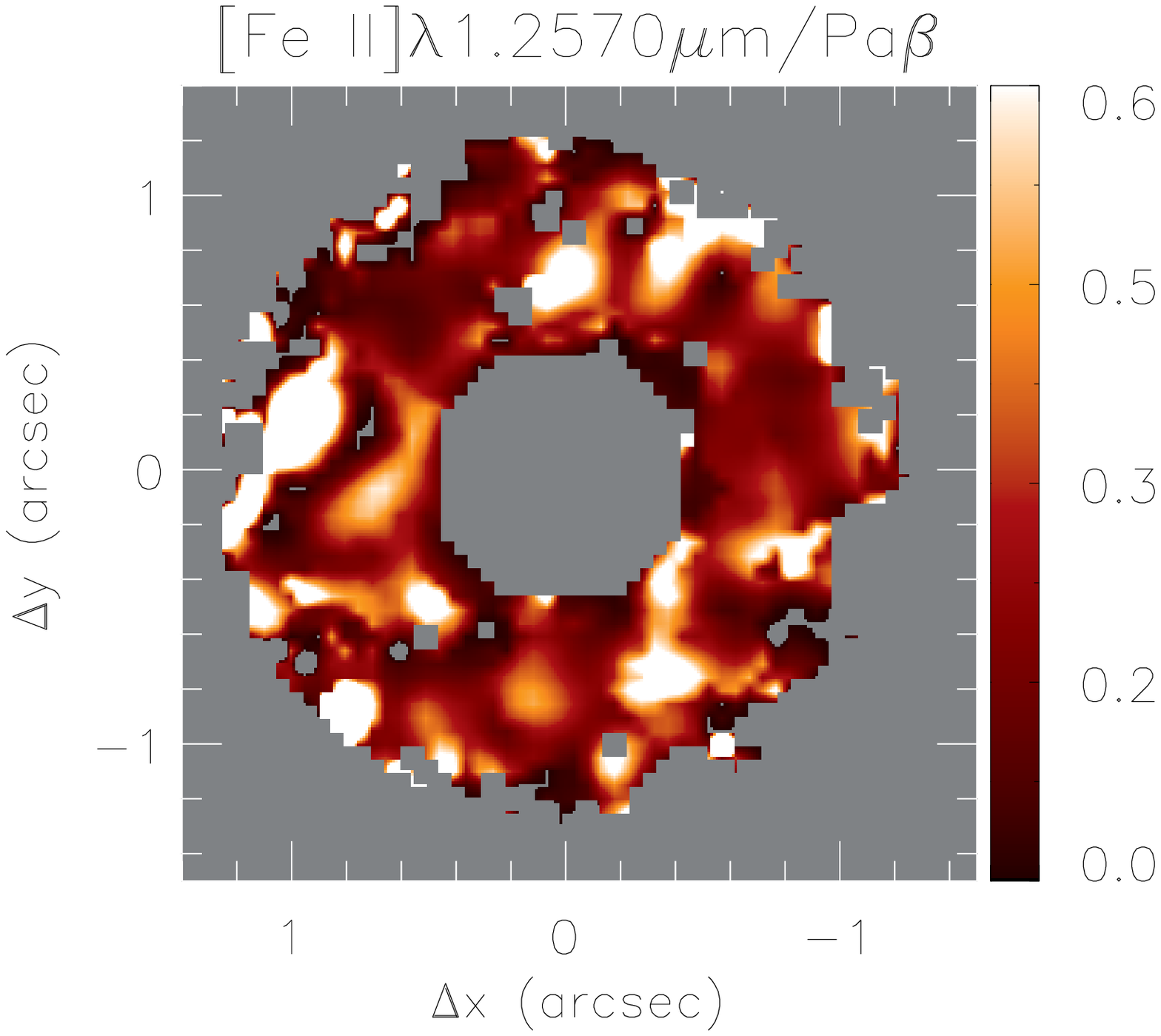}\quad   
    \includegraphics[height=4.8cm]{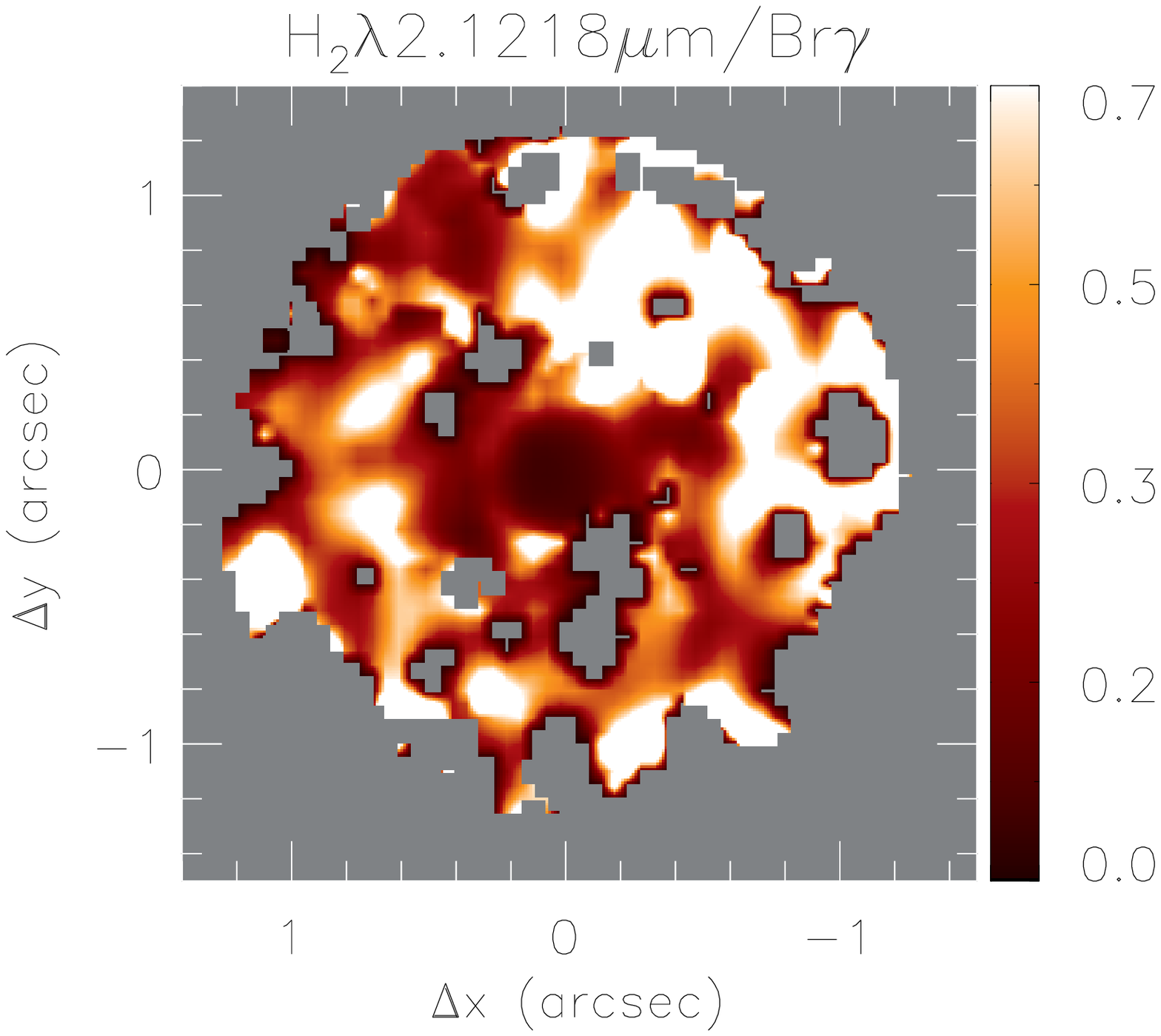}\quad
    \includegraphics[height=5.1cm]{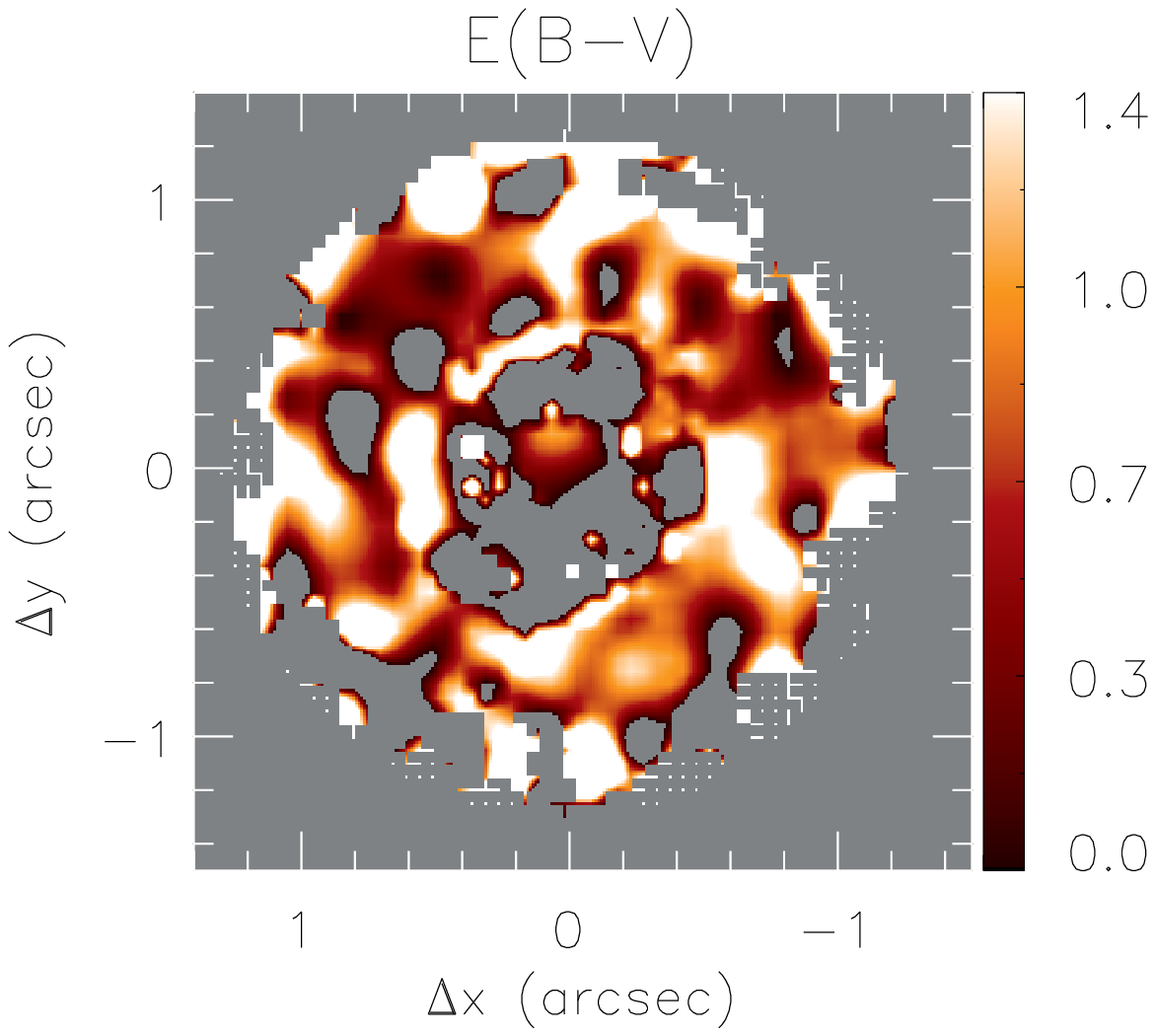}\quad

\caption{ [Fe\,{\sc ii}]$\,\lambda$\,1.25$\mu$m/Pa$\beta$ (left panel), H$\rm_2$\,$\lambda$\,2.12$\mu$m/Br$\rm gamma$ (middle panel) emission line ratio maps and reddening map obtained from the Pa$\beta$/Br$\rm gamma$ ratio (right panel). The gray regions represent locations where one or both lines do not present good flux measurements. }
\label{ratio}
\end{center}
\end{figure*}
\vspace{0.65cm}

The excitation mechanisms of the \feii\ and \hm\ emission lines can be investigated using the \fej/\pb\ and \hm/\br\ line ratios, respectively \citep{larkin98,reunanen02,ardila04,ardila05,rogerio2013,colina15,lamperti17}. We show maps of these line ratios for Mrk~42 in Fig.~\ref{ratio}. The left panel shows the \fej/\pb\ ratio map, while the \hm/\br\ map is shown in the middle panel.  Gray regions represent masked locations where the signal-to-noise ratio of one or both lines was not high enough to allow good fits. In addition, we do not show the values for the inner 0\farcs45, for which no \fej\ emission is detected (see Fig.~\ref{flux-ew}) and the \hm\ and \br\ are marginally detected and thus the uncertainty in the \hm/\br\ is high. In addition, the origin of the nuclear emission will be discussed in section \ref{ratios}.  


Both \fej/\pb\ and \hml/\br\ flux ratio maps show values smaller than 0.6 at most locations, suggesting that the \hm\ and \feii\ lines are excited by the star-forming regions \citep[e.g.][]{rogerio2013}. However, some differences are found between these maps. While \fej/\pb\ shows low values along the whole ring of CNSFRs, the \hml/\br\ shows higher values (of up to 0.7) for the CNSFRs located to the north-west of the nucleus.
 In Table~\ref{tab-ew} we present the \fej/\pb\ and \hml/\br\ line ratio values for the CNSFRs of Mrk~42, obtained by measuring the fluxes of the emission line from the integrated spectra shown in Fig.~\ref{spectra}, and identified in Fig.~\ref{flux-ew}. Small values of \fej/\pb\ ($<0.5$) and \hml/\br\ ($<1.0$) are observed for all CNSFRs.

\begin{table*}
\caption{Line-ratios, velocity dispersion and Equivalent Widths for the CNSFRs, as obtained from the integrated spectra of Fig.~\ref{spectra}. The values for the velocity dispersion were corrected for the instrumental broadening.}
 
\begin{tabular}{c c c c c c c c c c c} 
\hline
  &  \multicolumn{2}{c}{Ratios}           & \multicolumn{3}{c}{Velocity dispersion} & \multicolumn{4}{c}{Equivalent Widths}\\ \hline
P & \feii 1.25/\pb & \hm 2.12/\br &  \pb & \feii 1.64  & \hm 2.12 & \pb  & \feii 1.64   & \hm 2.12 & \br   \\ 
  &                &              &  (\kms)&(\kms)  & (\kms) & (\AA)  & (\AA)   & (\AA) & (\AA)   \\ 
\hline 
A &  0.18$\pm$0.06     & 0.38$\pm$0.10     & 25.0$\pm$1.3	&53.5$\pm$15.0    &35.8$\pm$9.2       &14.8$\pm$3.2   &2.9$\pm$0.8  &2.7$\pm$0.6  &7.4$\pm$1.2   \\ 
B &  0.47$\pm$0.05     & 0.61$\pm$0.18     & 29.8$\pm$2.8	&36.4$\pm$11.3    &40.5$\pm$11.3      &11.4$\pm$4.1   &2.4$\pm$0.7  &3.8$\pm$1.1  &6.6$\pm$1.7   \\ 
C &  0.18$\pm$0.07     & 0.62$\pm$0.17     & 26.8$\pm$1.7	&50.5$\pm$14.4    &40.5$\pm$10.6      &10.0$\pm$2.7   &3.0$\pm$0.8  &3.8$\pm$1.0  &6.6$\pm$1.7   \\
D &  0.22$\pm$0.08     & 0.59$\pm$0.16     & 21.4$\pm$1.7	&35.3$\pm$13.0    &35.2$\pm$8.9       &6.8 $\pm$1.8   &1.2$\pm$0.4  &2.2$\pm$0.5  &3.6$\pm$0.7   \\ 
E &  0.27$\pm$0.09     & 0.42$\pm$0.10     & 22.5$\pm$1.5	&36.1$\pm$11.3    &22.2$\pm$6.6       &8.8 $\pm$3.8   &1.7$\pm$0.5  &2.0$\pm$0.5  &4.9$\pm$0.8  \\ 
F &  0.20$\pm$0.05     & 0.95$\pm$0.23     & 31.4$\pm$2.3	&36.3$\pm$8.2	 &53.5$\pm$11.9       &10.9 $\pm$4.5   &2.8$\pm$0.7  &5.6$\pm$1.4  &5.8$\pm$0.9  \\ 
G &  0.21$\pm$0.09     & 0.72$\pm$0.15     & 29.7$\pm$2.3	&40.5$\pm$11.1    &25.8$\pm$5.8       &9.3 $\pm$2.5   &2.6$\pm$0.7  &3.0$\pm$0.6  &4.3$\pm$0.8  \\ 
H &  0.41$\pm$0.18     & 0.89$\pm$0.22     & 26.8$\pm$2.5	&34.3$\pm$12.2    &37.0$\pm$6.2       &5.8 $\pm$2.0   &1.4$\pm$0.4  &3.1$\pm$0.5  &3.5$\pm$0.8 \\ 
 \hline 
\end{tabular}
\label{tab-ew}
\end{table*}

The \pb/\br\ flux ratio can be used to map the gas extinction. The reddening ($E(B-V)$) can be obtained by

\begin{equation}
E(B-V)=4.74\,{log}\left(\frac{5.86}{F_{\rm Pa\beta}/F_{\rm Br\gamma}}\right),
\end{equation}

where $F_{\rm Pa\beta}$ and $F_{\rm Br\gamma}$ are the fluxes of $Pa\beta$ and $Br\gamma$ emission lines, respectively. We adopted the intrinsic ratio $F_{\rm Pa\beta}/F_{\rm Br\gamma}=5.86$ corresponding to case B recombination for an electron temperature $T_e=10^4$ K and electron density $N_e=100\,{\rm cm^{-3}}$ \citep{osterbrock06} and used the extinction 
law of \citet{cardelli89}. The resulting $E(B-V)$ map is shown in the right panel of Fig.~\ref{ratio}, where values ranging from 0 to 1.2 are derived. The last column of Tab.~\ref{tab-flux} shows the resulting $E(B-V)$ for the nucleus and for each CNSFR.  The reddening derived for the CNSFRs of Mrk\,42 are consistent with those obtained for star forming regions in other galaxies and H\,{\sc ii} galaxies \citep{Kotilainen,martins13,Riffel2016}.

\subsection{Gas velocity fields and velocity dispersion maps}

\begin{figure*}
\includegraphics[height=5.1cm]{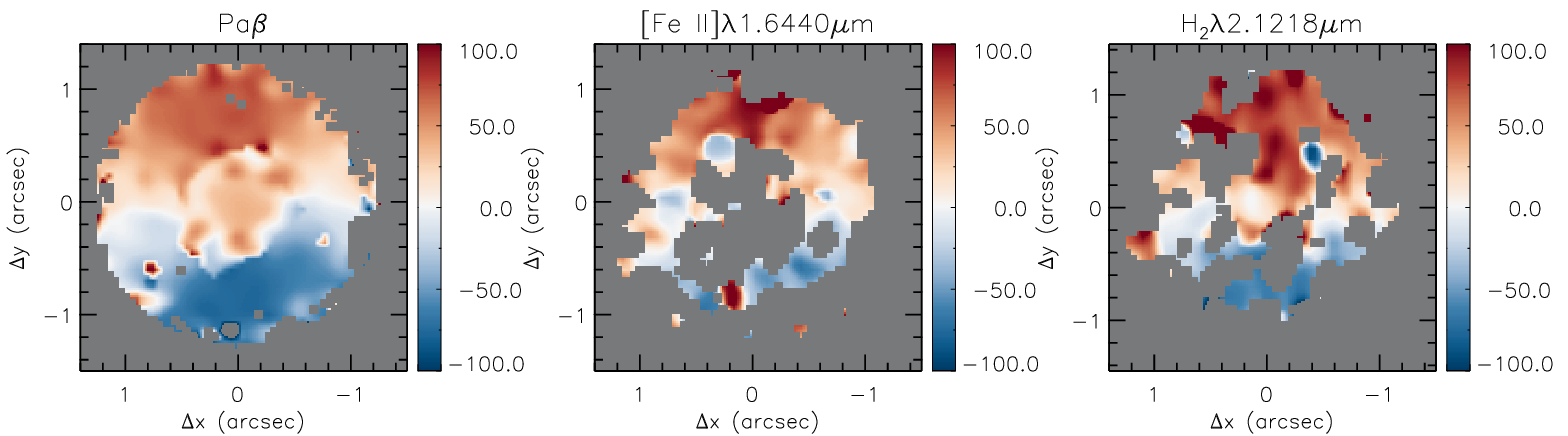}\quad  
\includegraphics[height=5.1cm]{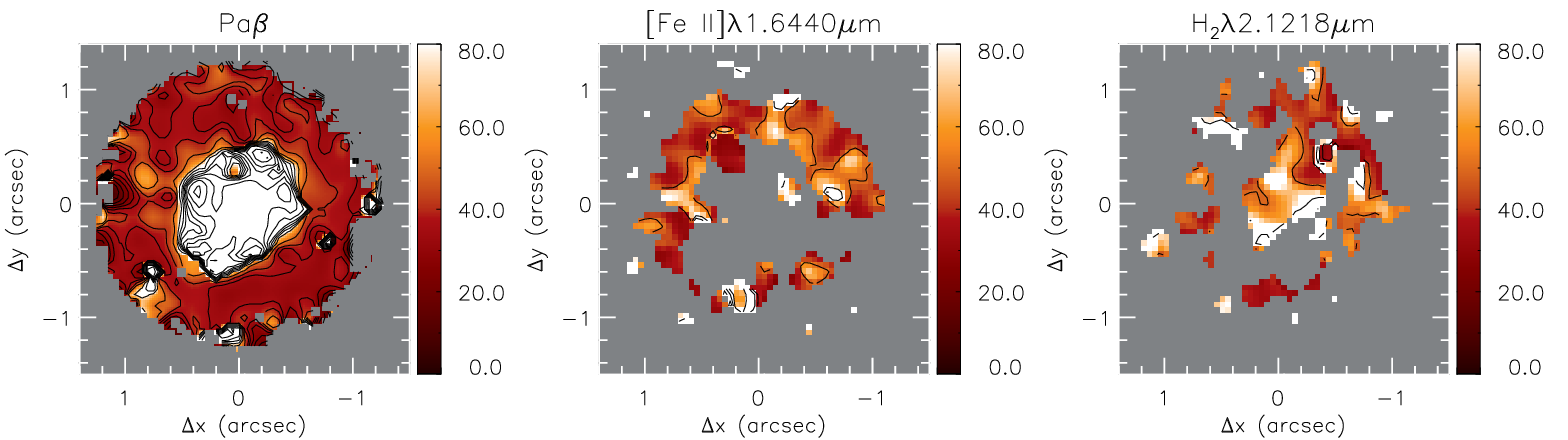}\quad  
\caption{Gas velocity fields (top) and velocity dispersion $\sigma$ maps (bottom) for the Pa$\beta$ (left), [Fe\,{\sc ii}]$\,\lambda$\,1.64$\mu$m (center) and H$\rm_2$\,$\lambda$\,2.12$\mu$m (right) emission lines. The color bars are shown in units of $\rm km\,s^{-1}$.}
\label{gas-kin}
\end{figure*}

Gas velocity fields (top panels) and velocity dispersion (bottom panels) maps are shown in Fig.~\ref{gas-kin}. Gray regions in the maps correspond to masked locations, using the same criteria as for the emission-line flux maps (see sect.~\ref{results}). The velocity fields for  \pb\ (top-left panel), \feii\ (top-center panel) and \hm\ (top-right panel) are presented after the subtraction of the systemic velocity of the galaxy ($V_{s} = 7390$\,\kms), as derived by the modeling of the \pb\ velocity field by a rotating disk (see Sec.~\ref{disc-kin}). At the inner 0\farcs45, where the \pb\ profile was fitted by two Gaussian curves, we use the results for the narrow component. The velocity fields for all lines are similar and suggest gas rotation in the plane of the galaxy with the line of nodes oriented approximately along the north-south direction and a line-of-sight velocity ($V_{\rm LOS}$) amplitude of 150\,\kms.  Besides ordered rotation, in the inner 0\farcs45 an additional velocity component is observed as indicated by the larger redshifts seen in \pb\ and \hm\ $V_{LOS}$ maps, possibly due to outflows from the nucleus.

The bottom panels of Fig.~\ref{gas-kin} show the velocity dispersion maps for the \pb\ (left panel), \feii\ (center panel) and \hm\ (right panel)  emitting gas. These maps were corrected for the instrumental broadening, as $\sigma=\sqrt{\sigma_{obs}^2 - \sigma_{ins}^2}$, where $\sigma_{obs}$ is the observed velocity dispersion and $\sigma_{ins}$ is the instrumental velocity dispersion given by $\sigma_{inst}={\rm FWHM/2.355}$ with the FWHM measured from typical lines present on the wavelength calibration lamp (as presented in Sec.~\ref{obs}).  Small values of $\sigma$ (<\,50\,\kms) are observed along the ring of CNSFRs for all emission lines, while higher values are seen within the inner 0\farcs45 for the \pb\ emission with $\sigma$ values of up to 200\,\kms. For illustration purpose,  the $\sigma$ map for \pb\ shows only values smaller than 80~\kms, which allows the comparison with the \hm\ and \feii\ $\sigma$ maps.  The high $\sigma$ values seen at the nuclear region for \pb\ support the interpretation that an outflow is present within the inner 0\farcs5. 
Some higher $\sigma$ values are also seen for the \hm\ at the nucleus, reaching values of up to 80~\kms. 

\begin{figure*}
\begin{center}
 \includegraphics[height=4.5cm,width=4.35cm]{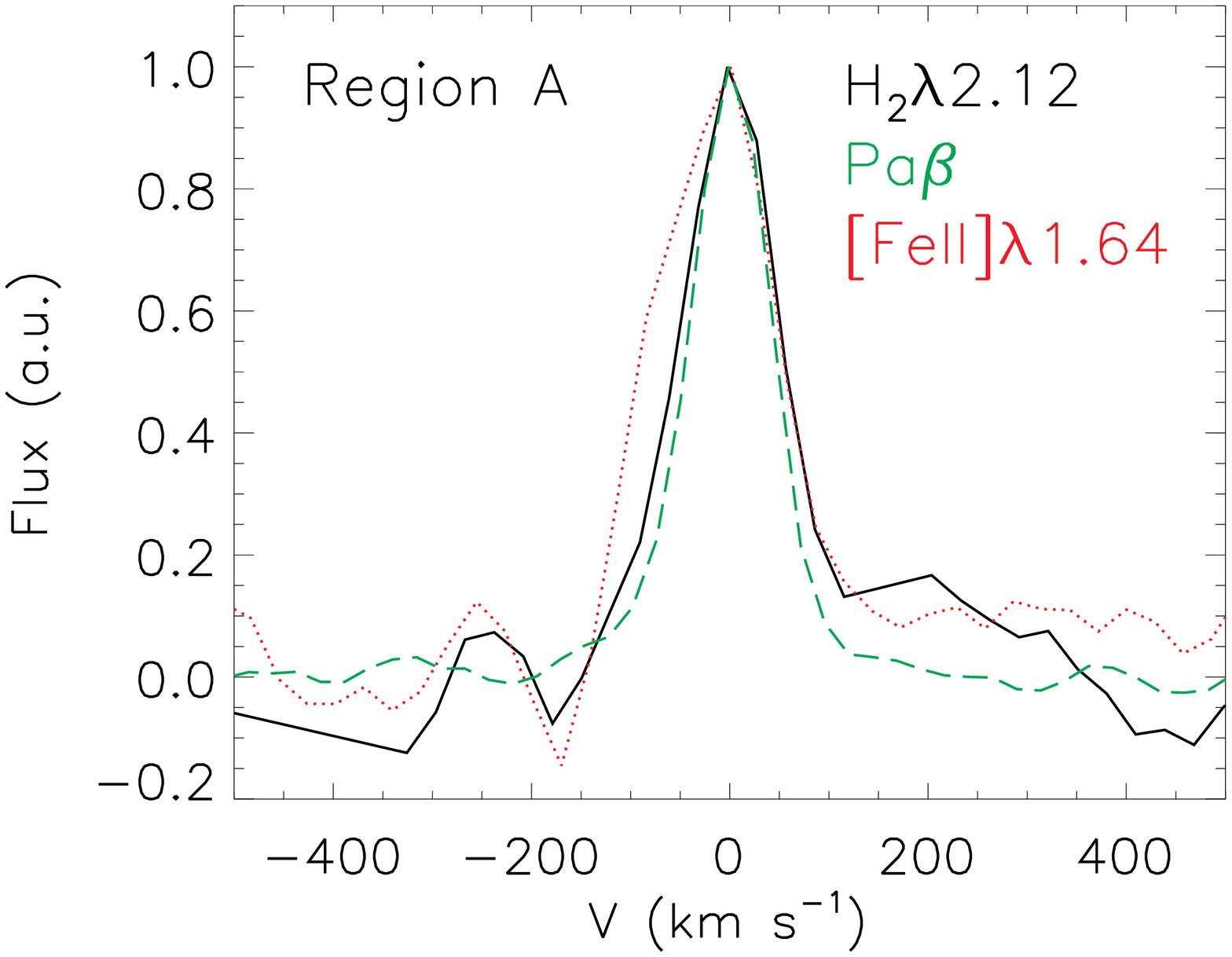}
  \includegraphics[height=4.5cm,width=4.35cm]{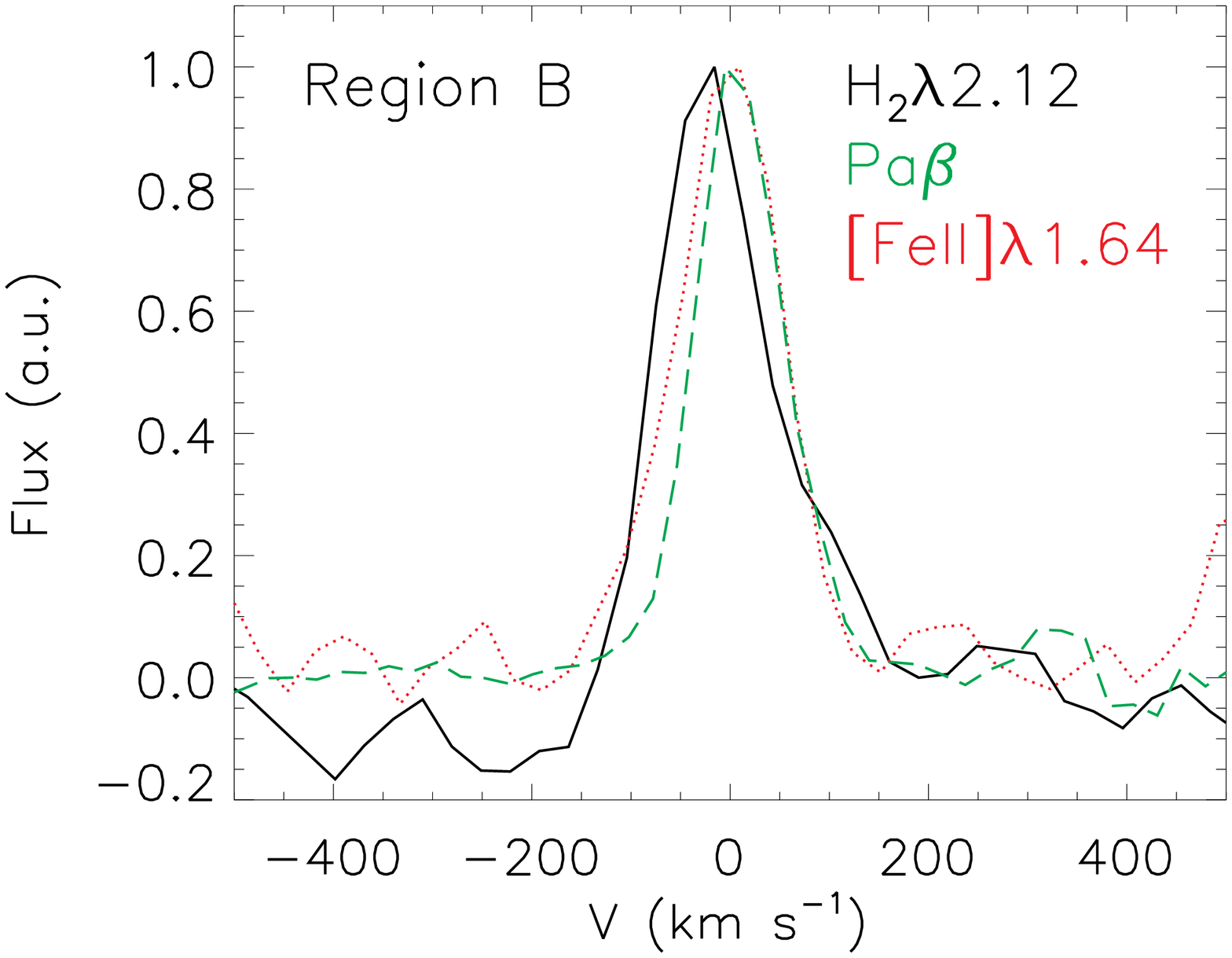} 
   \includegraphics[height=4.5cm,width=4.35cm]{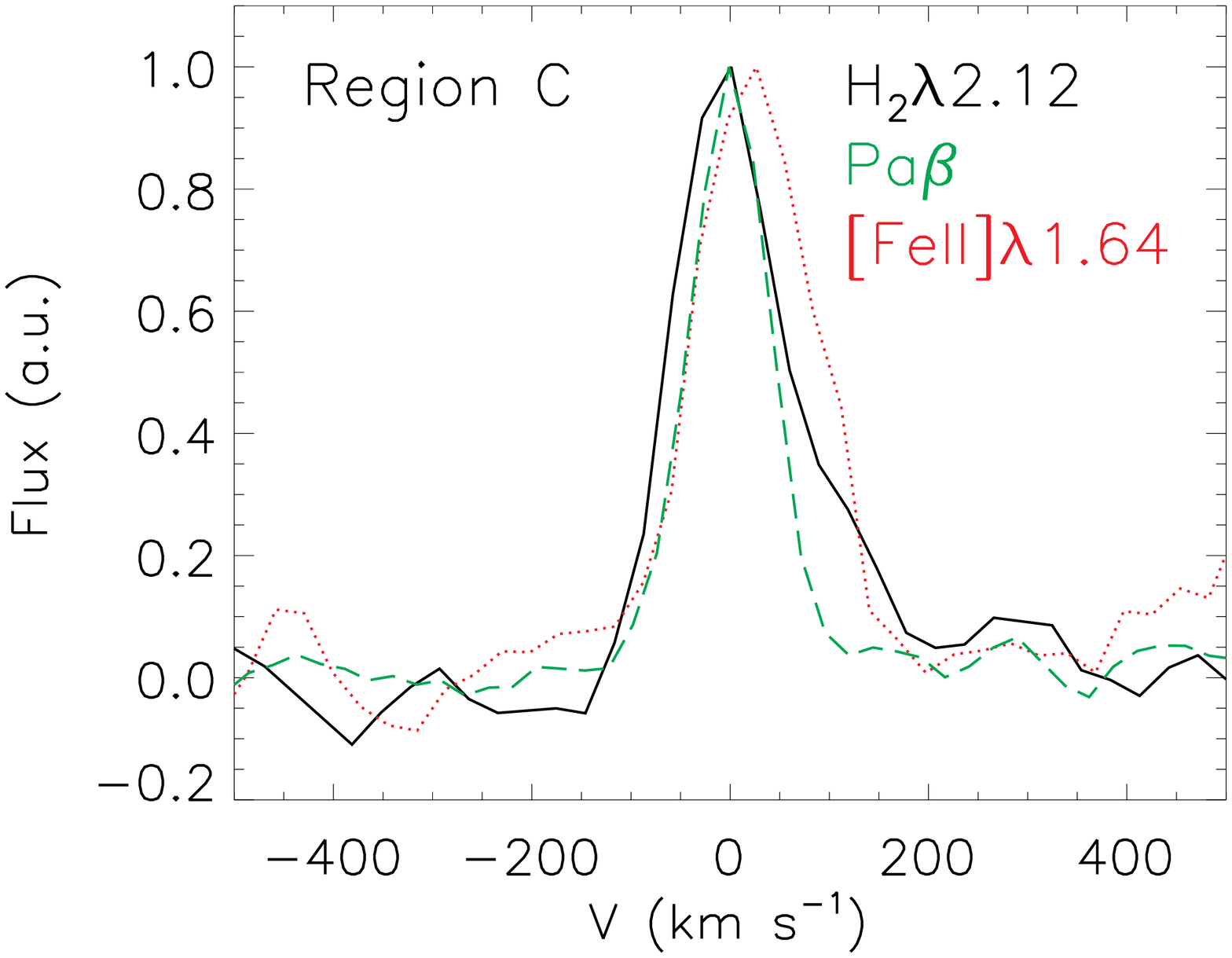} 
   \includegraphics[height=4.5cm,width=4.35cm]{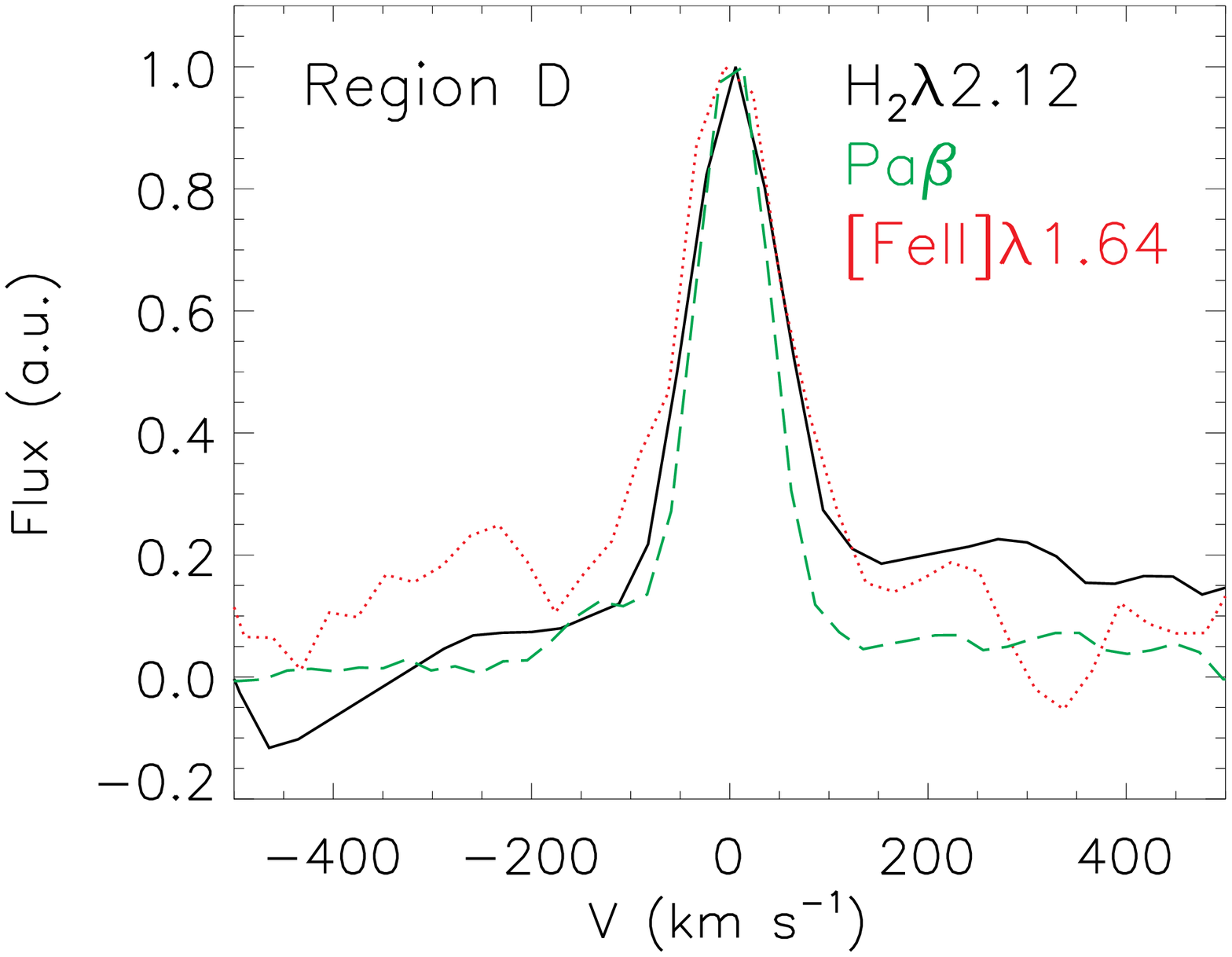} 

\vspace{0.3cm}

 \includegraphics[height=4.5cm,width=4.35cm]{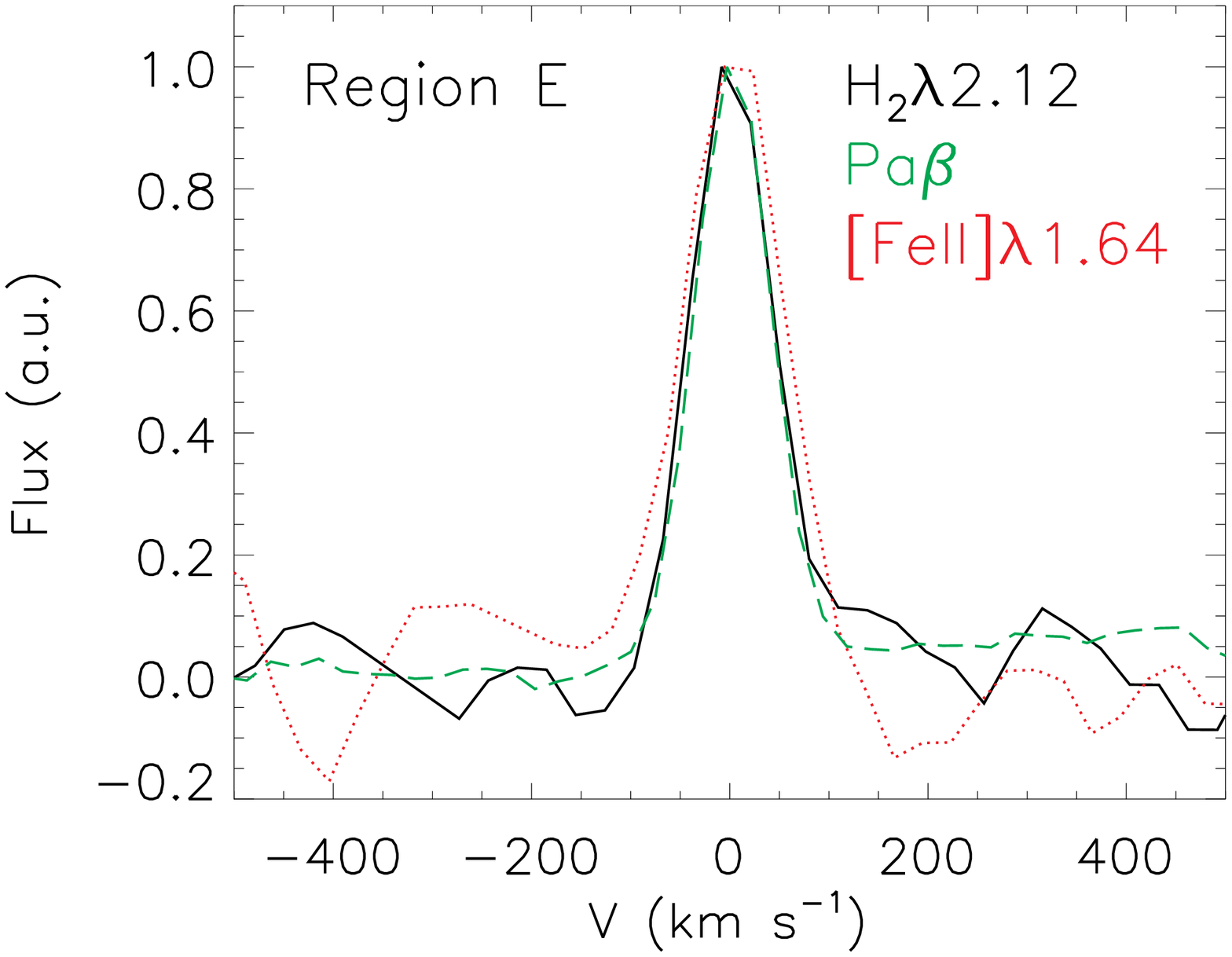}
  \includegraphics[height=4.5cm,width=4.35cm]{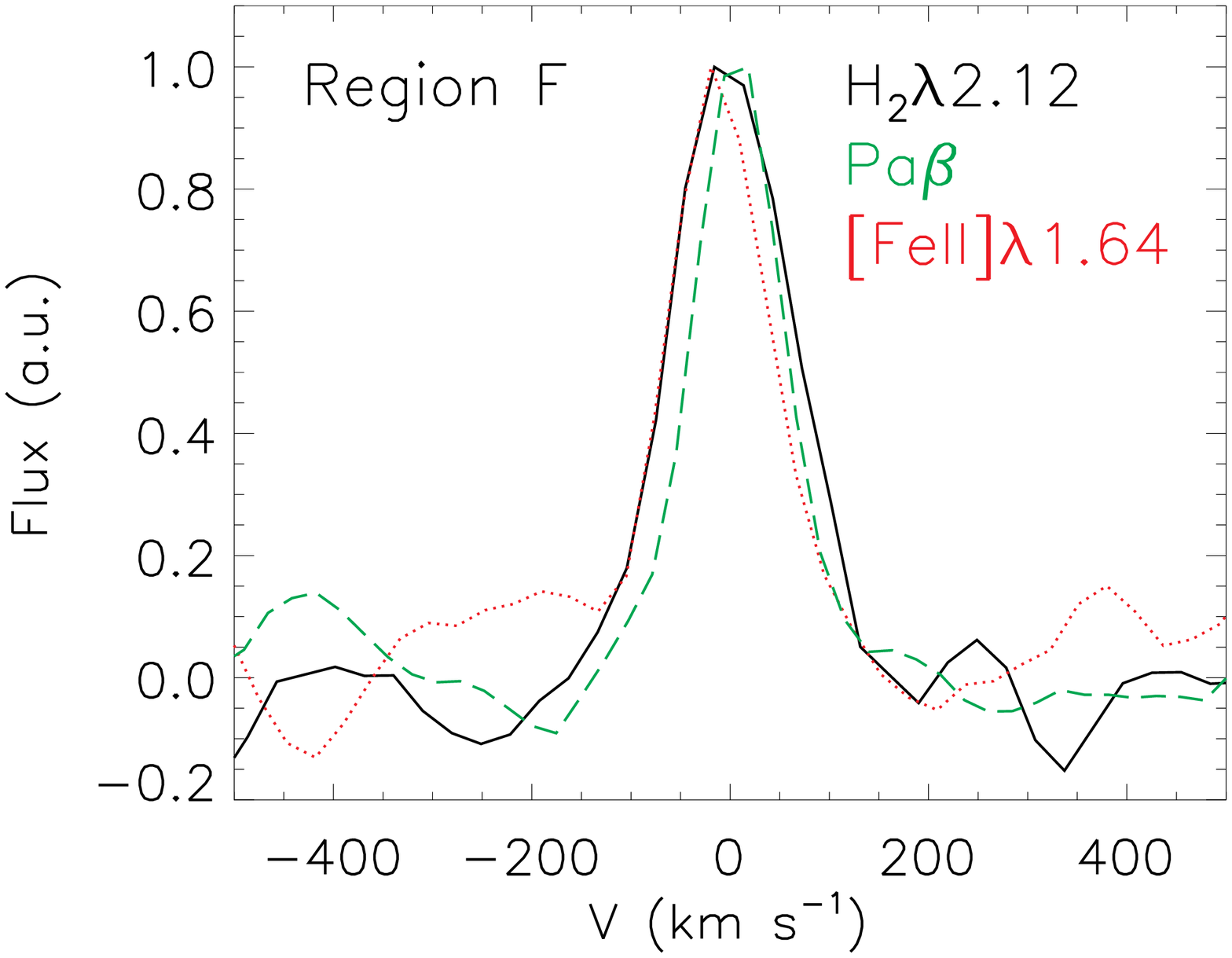} 
   \includegraphics[height=4.5cm,width=4.35cm]{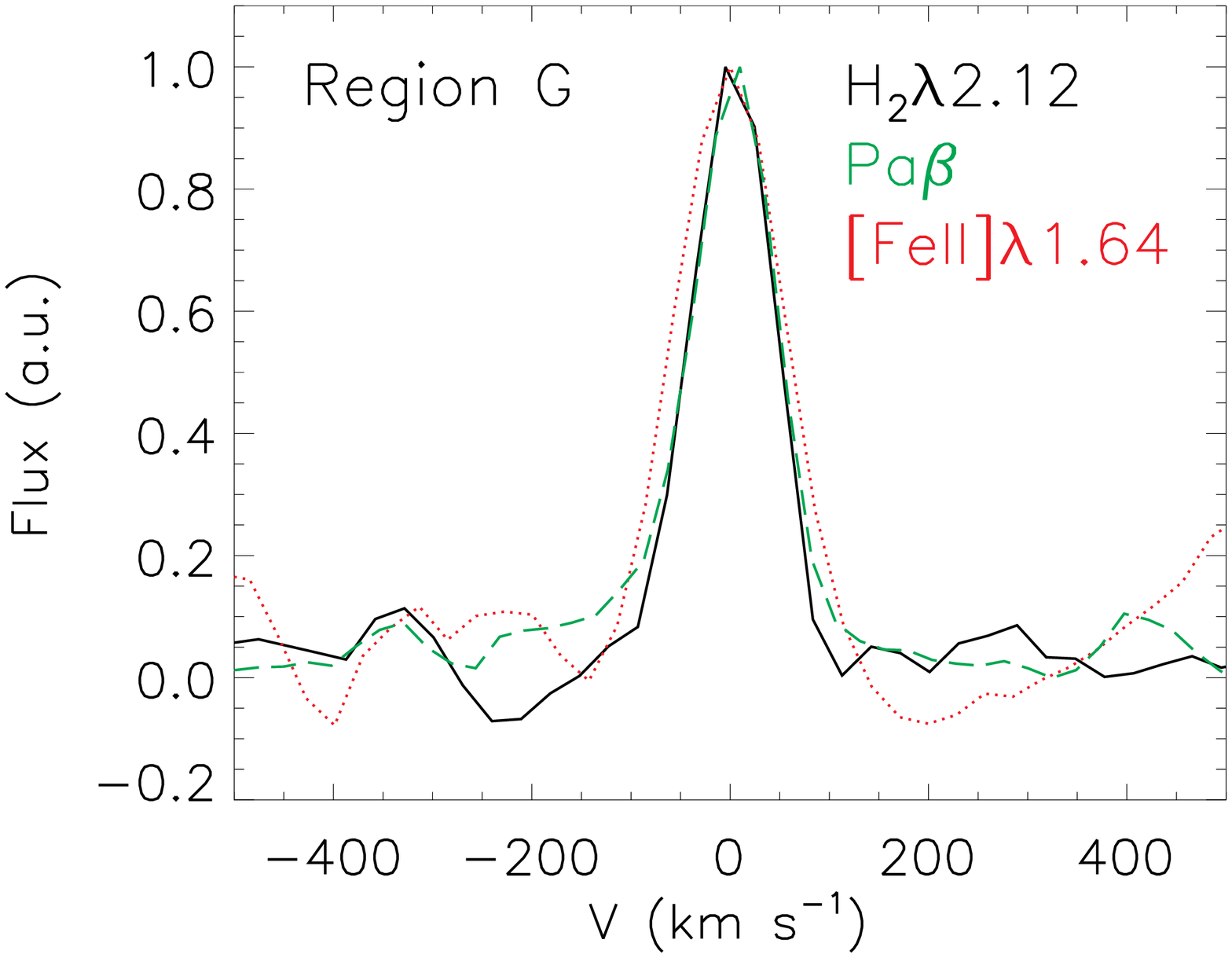} 
   \includegraphics[height=4.5cm,width=4.35cm]{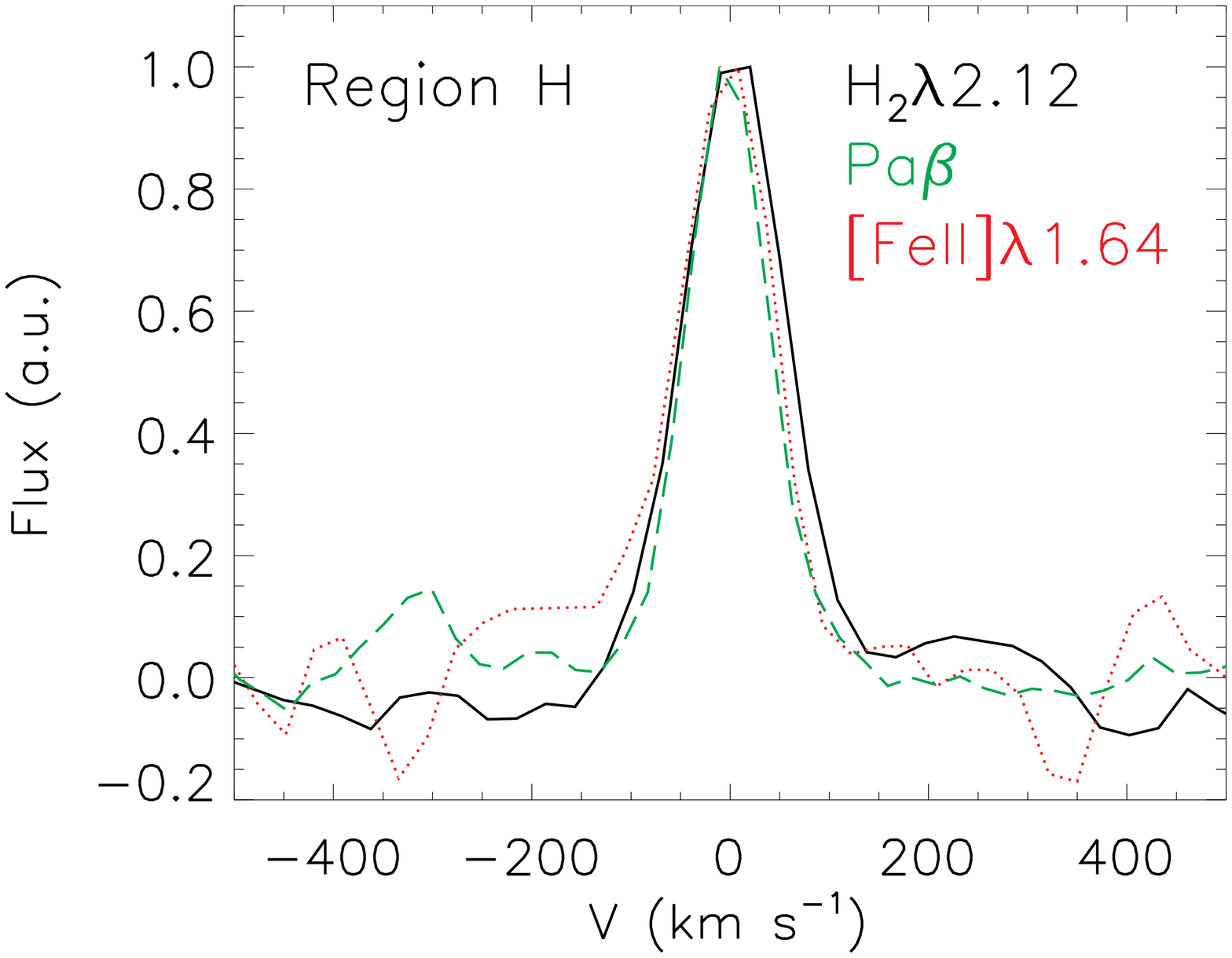} 
 \caption{Emission-line profiles of \hml\ (black), \pb\ (green) and \feh\ (red) for the CNSFRs. The panels show the normalized flux vs. velocity in units of $\rm km\,s^{-1}$ for each CNSFR.}
\label{profiles}
\end{center}
\end{figure*}

In Table~\ref{tab-ew} we present the velocity dispersion values obtained for each CNSFR by fitting the line profiles seen in the integrated spectra of Fig.~\ref{spectra}, which suggest that the \pb\ profile is usually narrower than the \hm\ and \feii\ profiles. In Figure~\ref{profiles} we over-plot the \pb, \feii\ and \hm\ line profiles from each CNSFR, which suggests that indeed the \hm\ and \feii\ profiles are broader than that of \pb\ for most regions, although it should be noticed that the \pb\ is only marginally resolved in our spectra.

\section{DISCUSSION}\label{discussion}

\subsection{Origin of the H$\rm_2$ and [Fe\,{\sc ii}] emission from the ring of CNSFRs}

The [Fe\,{\sc ii}]$\,\lambda$\,1.25$\mu$m/Pa$\beta$ line ratio can be used to investigate the excitation mechanism of the [Fe\,{\sc ii}] emission. Some studies indicate that the [Fe\,{\sc ii}] emission in active galaxies can be produced by shocks due to the interaction of radio jets with the interstellar medium \citep{forbes93,Dopita1996},  while other studies indicate that photo-ionization can produce most of the observed [Fe\,{\sc ii}] emission \citep{simpson96,dors12}. A way to disentangle what are the main excitation mechanisms of the \feii\ emission is based on the use of [Fe\,{\sc ii}]$\,\lambda$\,1.25$\mu$m/Pa$\beta$ flux ratio \citep{reunanen02,ardila04,ardila05,rogerio2013,colina15}. Typical values for Seyfert nuclei are 0.6$\:<\:$[FeII]1.2570$\mu$m/Pa$\rm\beta$$\:<\:$2.0, while smaller values than 0.6 are observed for Starburst galaxies and H\,{\sc ii} regions and values higher than 2.0 may be associated to shocks, e.g. due to radio jets or supernovae explosions \citep[e.g.][]{ardila04}. As observed in the left panel of Fig.~\ref{ratio}, the [FeII]1.2570$\mu$m/Pa$\rm\beta$ flux ratio map presents values smaller than 2.0 at all locations of the ring of CNSFRs, indicating that the origin of the \feii\ line emission is photo-ionization by young stars.

The origin of the H$\rm_2$ emission in active galaxies has been addressed by several studies \citep[e.g.][]{reunanen02,ardila04,ardila05,davies05,ra09,eso428,mrk1066a,rogerio2013,dors12,colina15}. In summary, 
the H$_2$ lines can be originated by two mechanisms: (i) fluorescent
excitation through absorption of soft-UV photons, e.g. from H\,{\sc ii} regions or AGNs \citep{black87} and (ii) collisional excitation due to heating of the gas 
by shocks \citep{hollenbach89} or by X-rays from the AGN  \citep{maloney96}. The H$\rm_2$\,$\lambda$\,2.12$\mu$m/Br$\rm\gamma$ line ratio can be used to investigate the H$\rm_2$ main excitation mechanism: values smaller than 0.6 are commonly observed in H\,{\sc ii} regions and Starburst galaxies, where the H$\rm_2$ lines are originated by fluorescence, while larger values are observed for AGNs and supernova remnants, where heating of the gas by X-rays and shocks may dominate \citep[e.g.][]{ardila05,dors12,rogerio2013,colina15}. For Mrk~42, the H$\rm_2$\,$\lambda$\,2.12$\mu$m/Br$\rm\gamma$  flux ratio map (middle panel of Fig.~\ref{ratio}) show values smaller than 0.6 at all locations of the CNSFRs ring, indicating that the observed H$_2$ emission is produced by fluorescent excitation due to absorptions of UV photons emitted by young stars.

\subsection{A nuclear Starburst in Mrk\,42}

Using the flux values presented in Table~\ref{tab-flux} for the nucleus of Mrk\,42 we obtain very low  H$\rm_2$\,$\lambda$\,2.12$\mu$m/Br$\rm\gamma$ and [Fe\,{\sc ii}]$\,\lambda$\,1.25$\mu$m/Pa$\beta$ values. Both ratios show values smaller than $\sim$0.1, which are not consistent with those usually observed for Seyfert nuclei \citep[larger than 0.6 for both ratios,][]{ardila04}.

 By fitting the SDSS \citep{albareti17,blanton17} spectrum of Mrk~\,42, we obtain [N\,{\sc ii}]$\lambda$6583/H$\alpha\approx 0.2$ and [O\,{\sc iii}]$\lambda$5007/H$\beta\approx3$, which put the nucleus of Mrk\,42 in the Starburst region of the BPT diagram \citep{bpt81}, consistent near-IR line ratios, although it should be noticed that the SDSS spectrum is obtained for an aperture of 3$^{\prime\prime}$ diameter, which includes the CNSFRs ring. However, it is well know that this galaxy presents an type 1 AGN, as broad recombination lines are detected. The low-emission line ratios detected at optical and near-IR can be explained if, besides the central AGN, Mrk~42 presents also strong nuclear Starburst activity, which dominates the gas excitation. Similar nuclear Starbursts are seen in other galaxies with CNSFRs rings, as for NGC\,4303 \citep{colina99,colina02,Riffel2016}.

\subsection{Gas kinematics}\label{disc-kin}

The velocity fields presented in Fig.~\ref{gas-kin} clearly show a rotation component. In order to obtain the kinematical and geometrical parameters, we have fitted the observed \pb\ velocities by a rotating disk model, assuming circular orbits in the plane of the galaxy, using the following equation \citep{bertola}:

\begin{equation}
\begin{multlined} 
V_{r}(R,\Psi) = V_{s} +  \\
 \frac{A\:R\:\rm cos(\Psi-\Psi_{0})\rm sin(\it{i})\rm cos^{\it p}\theta}{R^{2}[\rm sin^{2}(\Psi-\Psi_{0}) + \rm cos^{2}(\it{i})\rm cos^{2}(\Psi-\Psi_{0})] + c_{0}^{2}\rm cos^{2}(\it{i})^{\it p/2}},
\label{model-bertola}
\end{multlined}
\end{equation}
where ${V_{r}}$\, is the circular velocity and ${R}$ and ${\Psi}$ are the coordinates of each pixel in the plane of the sky, $V_s$ is the systemic velocity, \textit{A} is the amplitude of the rotation curve in the plane of the galaxy, $\Psi_{0}$ is the position angle of the line of nodes, \textit{c$_0$} is a concentration parameter, defined as the radius where the rotation curve reaches 70\,\% of the velocity amplitude, \textit{i} is the disc inclination relative to the plane of the sky and \textit{p} is a model fitting parameter. The parameter {\it p} measures the slope of the rotation curve where it flattens, in the outer region of the galaxy
and it is limited between 1 $\le$ {\it p} $\le$ 3/2. For {\it p}\,=\,1 the rotation curve at large radii is asymptotically flat
while for {\it p}\,=\,3/2 the system has a finite mass.

We fitted the \pb\ velocity field by the equation above using the MPFITFUN routine \citep{mark09} to perform a non-linear least-squares fit, where initial guesses are given for each parameter and the routine returns their values for the best fitted model. The \pb\ velocity field was chosen because the \pb\ emission line presents the highest signal-to-noise ratio among the observed emission lines. During the fit, we excluded regions at $r<$0\farcs45 from the nucleus, where an additional kinematic component is observed.  During the fit, the position of the kinematical center was kept fixed at the nucleus, defined as the location of the peak of the continuum emission.

\begin{figure*}
    \includegraphics[scale=0.56]{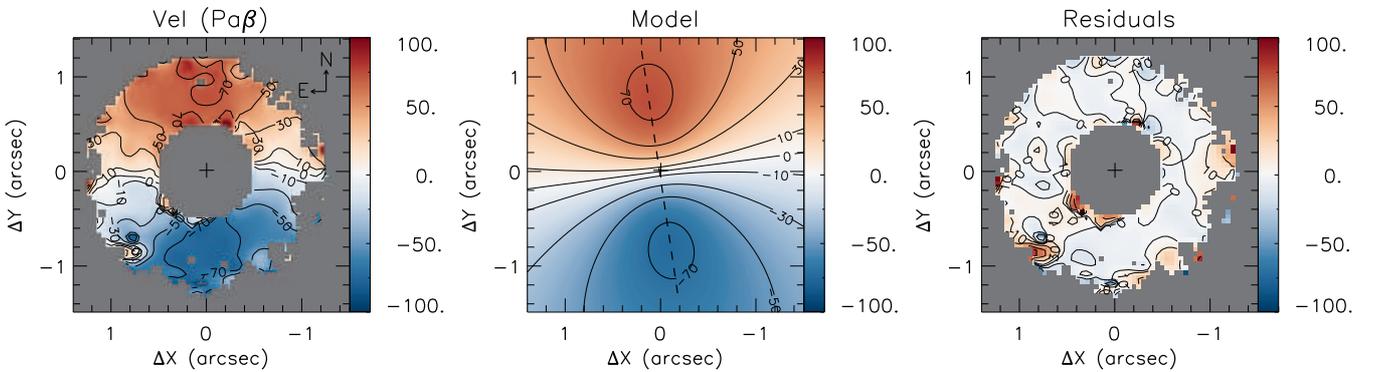}
\caption{Left panel:  Pa$\beta$ velocity field; middle: rotating disk model and right: residual map, obtained as the difference of the observed velocities and the model. The color bars show the velocity in units of km\,s$^{-1}$ and the central cross indicates the position of the nucleus. The external gray regions represent masked locations where the signal-to-noise ratio of \pb\ was not high enough to allow a good fit of its profile, while the gray region at $r<$0\farcs45 represents masked locations, where gas outflows are suggested. The central cross marks the position of the nucleus.}
\label{modelo_bertola}
\end{figure*}

The resulting kinematical model is shown in the middle panel of Fig.~\ref{modelo_bertola}, while the observed \pb\ velocity field is shown in the left panel. Gray locations represent masked positions, where we were not able to obtain good velocity measurements (distances larger than 1\farcs\,from the nucleus) and the nuclear region, where outflows from the central AGN seem to co-exist with a disk component. The residual map, obtained as the difference between the observed velocities and the model is presented in the right panel. The residual map presents values very close to zero at all locations (smaller than 10 \kms), showing that the velocity field is well represented by the model and thus, dominated by regular rotation.

The resulting parameters for the best fit model are: \textit{i} = 20.2$^{\circ}$$\pm$2.0$^{\circ}$, V$_s$ = 7391$\pm$11\,km\,s$^{-1}$, corrected to the heliocentric reference frame, \textit{A} = 259$\pm$24\,km\,s$^{-1}$, 
$\Psi_{0}$ =8$^{\circ}\pm$1$^{\circ}$, c$_{0}$ = 0.56 $\pm$ 0.05 arcsec and \textit{p} = 1.5, which was limited between 1 and 1.5. The disk inclination is slightly larger than that of the large scale disk, obtained assuming $i={\rm acos}{b/a}$, where $a$ and $b$ are the semi-major and semi-minor axis of the galaxy and the value of the systemic velocity ($V_s=7385$\,\kms) is in a good agreement with that of \citet{falco1999}. The orientation of the line of nodes is similar to that of the large scale disk, quoted in the Hyperleda database \citep[$\Psi_0=-12^\circ$,][]{makarov14}. 

\begin{figure}
\begin{center}
       \includegraphics[height=6.5cm]{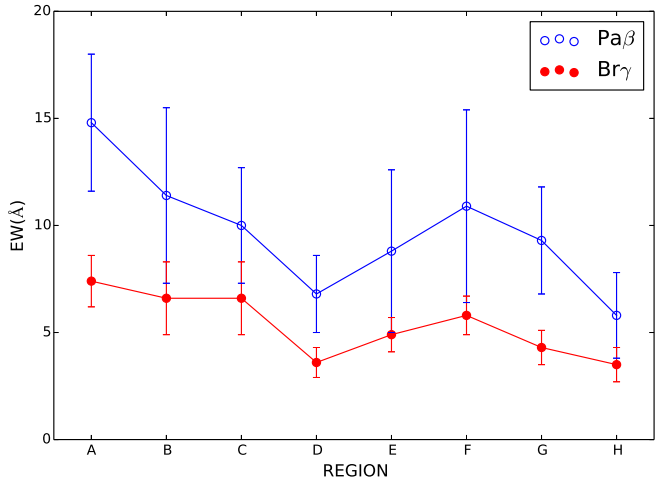}

\caption{Sequence of values  for the EqW of \pb\ (open blue circles) and \br\ (filled red circles) emission lines for the CNSFRs labeled from `A' to `H' in Fig.~\ref{flux-ew}. }
\label{plots}
\end{center}
\end{figure}

Within the inner  $r<$0\farcs45 radius, the \pb\ emission line seems to present two narrow kinematic components: the blueshifted component is seen with velocity of $V_{\rm blue}\approx-300$\,\kms, relative to the systemic velocity, while the redshifted component is seen with velocities very close to the nuclear velocity. The blueshifted component is only marginally detected and thus, we were not able to properly fit it. A possible interpretation of these two components is that the blueshifted one is originated from outflowing gas from the nucleus excited by the central AGN, while the redshifted one may be originated from the nuclear Starburst located at the disk of the galaxy.

\subsection{Stellar populations and CNSFRs}

We used the integrated emission-line fluxes of Table~\ref{tab-flux} to calculate the following properties: the mass of ionized M$_{\rm H\:II}$ gas; rate of ionizing photons $Q$[H$^{+}$]; the star formation rate (SFR) and the mass of hot molecular M$_{\rm H2}$.

The mass of ionized gas can be estimated by \citep[e.g.][]{Riffel2016}
 
\begin{equation}
\left(\frac{M_{\rm H\:II}}{M_{\rm \odot}}\right) = 3 \times 10^{19}\left(\frac{F_{\rm Br\gamma}}{\rm erg\:cm^{-2}\:s^{-1}}\right) \left(\frac{D}{\rm Mpc}\right)^{2} \left(\frac{N_{e}}{\rm cm^{-3}}\right)^{-1},
\label{eq-HII}
\end{equation}
 where F$_{Br\gamma}$\ is the Br$\gamma$ flux, \textit{D} is the distance to the galaxy and \textit{N$_{e}$} is the electron density, under the assumption of case B recombination at an electron temperature of  $10^{4}$K \citep{osterbrock06}. The mass of hot molecular gas can be estimated by \citep{scoville1982,n4051}

\begin{equation}
\left(\frac{M_{\rm H_2}}{M_{\odot}}\right) = 5.0776 \times 10^{13} \left(\frac{F_{\rm H_{2}\lambda\,2.1218}}{\rm erg\:cm^{-2}\:s^{-1}}\right) \left(\frac{D}{\rm Mpc}\right)^{2},
\label{eq-H2-quente}
\end{equation}
where $F_{H_{\rm 2}\lambda\,2.1218}$ is the flux of the H$\rm_2$\,$\lambda$\,2.1218 emission line and we assume local thermal equilibrium and an excitation temperature of 2000 K for the H$_{\rm 2}$.

The emission rate of ionizing photons is obtained following \citet{n7582}:

\begin{equation}
\left(\frac{Q\rm [H^{+}]}{\rm s^{-1}}\right) = 7.47 \times 10^{13} \left(\frac{L_{\rm Br\gamma}}{\rm erg\:s^{-1}}\right),
\label{eq-fotons-ionizantes}
\end{equation}
where L$_{\rm Br\gamma}$ is the Br$\gamma$ luminosity and the star formation rate is obtained by:

\begin{equation}
\left(\frac{SFR}{M_{\odot}\rm yr^{-1}}\right) = 8.2 \times 10^{-40} \left(\frac{L_{\rm Br\gamma}}{\rm erg\:s^{-1}}\right),
\label{eq-SFR}
\end{equation}
following \citet{kennicutt98}. The equations above for $SFR$ and  $Q[\rm H^{+}]$ are derived under the assumption of continuous star formation, and should be considered just proxies of these parameters.

In order to estimate the mass of ionized gas and the resulting physical properties of each CNSFR of Mrk~42 that are shown in Table~\ref{prop}, we have assumed the electron density value of $N_e=300$ \,cm$^{-3}$, which is the mean value of the electron density for a sample of CNSFRs derived from the [S\,{\sc ii}]$\lambda$\,6717/$\lambda$\,6731 intensity ratio by \citet{diaz07} and \citet{dors08}. 


The derived physical parameters for the CNSFRs of Mrk~42 can be compared with previously published values. \citet{dors08} used optical IFS to study the CNSFRs of the Seyfert 2 galaxies NGC\,1097 and NGC\,6951 and found $SFR$ in the range 0.002--0.14~M$_\odot$\,yr$^{-1}$ for the eight star-forming regions studied. \citet{shi06} used the observations from the Sloan Digital Sky Survey (SDSS) to study a sample of 385 objects classified as star-forming galaxies and found an average value for the $SFR$ of 0.14~M$_\odot$\,yr$^{-1}$. \citet{falcon2014} used SINFONI data to study the star formation in the inner 700 pc of the active  spiral galaxy NGC\,613 and obtained $SFR\sim\,10^{-2}-10^{-1}$~M$_\odot$\,yr$^{-1}$ for eight CNSFRs. \citet{Riffel2016} found $SFR$  of (0.4-2.0)$\times 10^{-2}$~M$_\odot$\,yr$^{-1}$ for the CNSFRs of NGC\,4303 derived from SINFONI observations.  Thus, the values of $SFR$ derived for the ring of Mrk~42 (0.07--0.2 M$_\odot$\,yr$^{-1}$) are within the range of values observed for other galaxies with similar nuclear rings.

The emission rates of ionizing photons for the CNSFRs of Mrk~42 are in the range log($Q$)$=51.8-52.3$ photons\,s$^{-1}$, in agreement with values already obtained for CNSFRs in other galaxies \citep[e.g.][]{galliano08,n7582,wold06,Riffel2016}. Such rates correspond to more than hundred O3 stars in each CNSFR \citep{osterbrock06}, but these values should be considered only as a proxy as a continuous star formation regime was assumed and the most probable scenario may be that of instantaneous star formation.

Finally, the masses of ionized and hot molecular gas observed for Mrk~42 are similar to that observed for other CNSFRs \citep[e.g.][]{Riffel2016,n7582}. The average ratio between the ionized and hot molecular gas masses is $<M_{\rm H\:II}/M_{\rm H_2}>\sim$\,3312, being similar to that observed for the CNSFRs of NGC\,4303 \citep[2030,][]{Riffel2016}, and also in the range of of values obtained for the inner kiloparsec of nearby Seyfert galaxies, of  200--8000 \citep[e.g.][]{n5548,riffel18}.

 The total mass of ionized gas in the ring (derived by measuring the fluxes of \br\ and \hml\ from an integrated spectrum comprising distances between 0\farcs45 and 1\farcs3 from the nucleus is 1.9 times larger than the sum of $M_{\rm H\:II}$ of all CNSFRs. The total mass of hot molecular gas in the ring is 2.8 times larger than that of the CNSFRs, indicating that a large amount of gas is observed outside the CNSFRs.

\begin{table*}
\caption{Physical parameters of the CNSFRs in Mrk 42 from integrated emission-line fluxes shown in Table~\ref{tab-flux}. The location of each region is indicated in Fig.~\ref{flux-ew}. $Q$[H$^{+}$] is the ionizing photons rate, SFR the star formation rate, M$_{\rm H\:II}$ is the mass of ionized gas and M$_{\rm H2}$ the mass of hot molecular gas. }
\begin{center}
\begin{tabular}{c c c c c c} 
\hline
Region & log $Q$[H$^{+}$] (s$^{-1}$) &  SFR (10$^{-2}$M${_\odot}$yr$^{-1}$) &  M$_{\rm H\:II}$ (10$^{4}$M${_\odot}$) &  M$_{\rm H2}$ (M$_{\odot}$) \\ 
\hline
A     & 52.3$\pm$0.06    &  21.5$\pm$3.30         &  21.9$\pm$3.36      &  43.0$\pm$9.20 \\ 
B     & 51.8$\pm$0.06    &  7.4$\pm$1.20          &  7.6$\pm$1.23       &  23.6$\pm$6.65 \\ 
C     & 51.8$\pm$0.06    &  7.4$\pm$1.17          &  7.5$\pm$1.19       &  23.6$\pm$5.30 \\ 
D     & 51.8$\pm$0.07    &  7.6$\pm$1.43          &  7.7$\pm$1.46       &  23.0$\pm$4.72 \\ 
E     & 51.9$\pm$0.05    &  8.2$\pm$0.93          &  8.4$\pm$0.95       &  17.9$\pm$3.67 \\
F     & 51.8$\pm$0.05    &  7.8$\pm$0.94          &  7.9$\pm$0.96       &  38.5$\pm$7.86 \\ 
G     & 51.8$\pm$0.05    &  6.9$\pm$0.89          &  7.0$\pm$0.91       &  25.7$\pm$4.25 \\ 
H     & 51.8$\pm$0.08    &  6.4$\pm$1.27          &  6.5$\pm$1.30       &  29.6$\pm$4.13 \\ 
Ring  & 53.1$\pm$0.04    & 138.7$\pm$12.89        & 141.43$\pm$13.13    & 624.3$\pm$87.19 \\
\hline 
\end{tabular}
\end{center}
\label{prop}
\end{table*}

We can also use the spectra shown in Fig.~\ref{spectra} to better investigate the stellar content of the CNSFRs.  These spectra show strong absorption features at 1.17-1.20\,$\mu$m, which may be molecular absorptions from stars in the thermally pulsing asymptotic giant branch (TP-AGB) phase, as suggested by simple stellar populations models \citep{maraston05} and detected in the nuclear spectra of nearby AGNs \citep[e.g. VO and CN features, ][]{rogerio07,rogerio15}. In addition, these features are clearly present in the spectra of stars in the TP-AGB phase \citep{lancon2002,maraston05,rayner09}. By comparing the spectra of the CNSFRs `E', `F' and `G', with the stellar spectra shown by \citet{lancon2002}, one can see that similar features are also seen in  the spectra of oxygen (e.g. from TiO and VO) and carbon (e.g. CN features) rich stars in the upper AGB phase. 
\citet{martins13} also reported the detection of signatures of stellar
population dominated by the TP-AGB (as CN, TiO and ZrO absorptions) in near-IR spectra of H\,{\sc ii} galaxies, which show a similar absorption features at $\sim$1.19\,$\mu$m as seen in the CNSFRs of Mrk\,42. However, stellar population synthesis performed for star-forming galaxies by  \citet{martins13b} do not find a clear correlation between the near-IR (e.g. CN) indexes and the presence of intermediate-age stellar populations, but this is attributed by these authors as due to observational limitations. Thus, the detection of these molecular features in the CNSFRs `E', `F' and `G' suggests that these regions are older than the other CNSFRs of Mrk\,42 and already present some evolved stars.


\subsection{How does the star formation proceed in the CNSFRs ring?} \label{scenario}

Gas inflows towards the central region of galaxies provide a gas reservoir, which can trigger (circum-)nuclear star formation and a central AGN.  
The way the material flows towards the central region may be through bars, spiral arms or interactions with neighbor galaxies. Inflows of gas associated to nuclear dust spirals and bars have been observed for nearby active galaxies at scales of a few hundred parsecs, mostly using optical and near-IR IFS \citep[e.g.][]{n4051,fathi06,sb07,ms09,vandeVen10,mrk79,sm14a,sm17}, but the presence of nuclear dust structures is not a sufficient condition to produce gas streaming \citep{brum17}.

Two possible scenarios have been proposed for the origin of the CNSFRs at scales of few hundred parsecs: the {\it popcorn} and {\it pearls on a string} scenarios \citep{Boker2008} that show distinct distributions in age of the stellar clusters.   In the \textit{popcorn} scenario, the gas is accumulated in the circumnuclear region resulting in a ring of higher density material, which allows a simultaneous star formation in the whole ring or in individual hotsposts, collapsing at different times within the ring. Thus, for this scenario there is no relation between the age of the recent formed stellar clusters and their locations within the ring.    

In the {\it pearls on a string} scenario, a short-lived, quasi-instantaneous burst of star formation is induced and located in a certain region of the ring. The locations of these  over-density regions (ODR, where the gas density is high enough to trigger the star formation) depends on details of the gravitational potential, but they are frequently found at locations where the gas reaches the central region, as for example, at the tips of bars. Stellar clusters are formed in these ODRs, which follow orbits within the ring, resulting in several stellar clusters  with a sequence of ages, and are called ``pearls on a string". Thus, in this scenario, two segments of stellar clusters, on opposite sides of the ring, are expected in which the youngest clusters have their locations near the ODRs, becoming older when approaching the opposite ODR, i.e. there is a correlation between the stellar cluster age and the ring localization.

In order to distinguish between the {\it popcorn} and {\it pearls on a string} scenarios for the origin of the ring of CNSFRs in Mrk~42, we constructed Fig.~\ref{plots} that shows the sequence of EqW values for \pb\ (open blue circles) and \br\ (filled red circles) emission lines for the regions labeled from 'A' to `H'. The EqW values were measured from the integrated spectra of Fig.~\ref{spectra} and are shown in Table~\ref{tab-ew}. The EqW values of H recombination lines can be used to determine the age of the star formation regions, with higher values observed for younger regions and lower values for older regions, under the assumption that all regions are intrinsically equal in stellar mass and the mass in the burst is the same  \citep{dottori1981,copetti1986}.
By comparing \br\ EqW values observed for the CNSFRs of Mrk~42 with the predicted values by evolutionary photo-ionization models \citep{n7582,dors08}, we conclude that all regions have ages between 5 and 6 Myr, considering the correction due to the contribution to the bulge stars to the underlying continuum in the derived EqW values, as discussed  in Sec.~\ref{FeEqW}. The highest EqW values are observed for region `A', decreasing to a minimum value for region `D'. A second gradient is seen from region `F' to `H'. As the EqW of the H recombination lines can be associated with the age of the stars, these gradients can be interpreted as a signature of two age sequences, one from region `A' to `D' and another from region `F' to `H', favoring thus the {\it pearls on a string} scenario if the gas reaches the center along the major axis of the galaxy (close to the orientation of the bar - see Fig.~\ref{flux-ew}). On the other hand, the spectra for the regions `E', `F' and `G' clearly show the presence of the molecular absorption bands at 1.17--1.20\,$\mu$m. These absorptions (TiO, CN or VO) are originated in more evolved stars that are in the TP-AGB phase \citep[e.g.][]{maraston05,lancon2002}, indicating that these regions are older than the remaining and suggests no clear sequence in age for these CNSFRs and thus not supporting the above sequence of ages for these CNSFRs.


\subsection{Mass of the SMBH} \label{bhmass}



The mass of the SMBH can be obtained directly from the \pb\ width and luminosity using the following equation \citep{kim2015,kim10,woo15}: 
\begin{equation}
\begin{multlined}
\frac{M_{BH}}{M_{\rm \odot}} = 10^{7.04\pm0.02} \left(\frac{L_{\rm Pa\beta}}{10^{42} {\rm erg s^{-1}}}\right)^{0.48 \pm {0.03}}\, \left(\frac{\rm FWHM_{\rm Pa\beta}}{10^{3} {\rm km s^{-1}}}\right)^{2},
\label{eq-MBNS-kim}
\end{multlined}
\end{equation}
where $L_{\rm Pa\beta}$ is the luminosity  and  FWHM$_{\rm Pa\beta}$ corresponds to width of the broad ${\rm Pa\beta}$  component. Assuming $L_{\rm Pa\beta}=(9.0\pm 0.3)\times 10^{39}$\,erg\,s$^{-1}$ and ${\rm FWHM_{\rm Pa\beta}}=1480\pm$40\,\kms, measured from the \pb\ profile within a circular aperture of 0\farcs25 radius,  we obtain $M_{\rm BH} = 2.5^{+0.7}_{-0.6}\times10^{6}$~M$_{\odot}$.

The mass of the SMBH of Mrk~42 derived here is in good agreement with values obtained using scaling relations based on optical observations and  based on the X-ray excess variance, which are in the range (0.7-18)$\times$10$^{6}$ ~M$_{\odot}$ \citep{wang01,bian03,nikolajuk09}.

We notice that the width of the \pb\ broad emission line is larger by a factor of 1.7--2.2 than those quoted in the literature for the H$\beta$ broad component \citep{goodrich89,bian03,wang01}. However, it should be noted that the values of FWHM for H$\beta$ available in the literature are based on more than 20 years old low resolution spectra. Using the Sloan Digital Sky Survey \citep[SDSS,][]{albareti17,blanton17} spectrum of Mrk~\,42, we note that the broad H$\beta$ component is just marginally detected, but its width is consistent with our measurement for \pb. The H$\alpha$ broad component of Mrk\,42 is clearly detected in the SDSS spectrum and has a FWHM$\approx$1400\,\kms, being similar to that derived for \pb.


\section{CONCLUSIONS}\label{conclusions}

We used J, H and K bands Gemini NIFS observations to map the flux distributions and kinematics in the near-IR emission lines of the inner $1.5\times1.5$ kpc$^2$ of Mrk~42 at a spatial resolution of 60\,pc and velocity resolution of $\sim$40\,\kms. The main conclusions of this work are:

\begin{itemize}

\item  The emission-line flux distributions and EqW maps clearly show the ring of CNSFRs, previously detected in HST image at $\sim300$\,pc from the nucleus. 
We detected 8 CNSFRs (labeled from `A' to `H') in the ring and by comparing the observed EqW values of the Br$\gamma$ emission line with those predicted by photo-ionization models, we conclude that their age is between 5 and 6 Myr. 
These regions present star formation rates in the range 0.06--0.21 M$_\odot$\,yr$^{-1}$, ionizing photons rate of log($Q$)$=51.8-52.3$ photons\,s$^{-1}$, masses of ionized gas in the range (6.5--22)$\times$10$^{4}$~M${_\odot}$ and masses of hot molecular gas between 18 and 43 M$_{\odot}$. These masses represent $\approx$\,53\% and 36\% of the total masses in the ring, respectively. 

\item In addition to a central AGN, line-intensity ratios indicate that Mrk~42 harbors a nuclear Starburst. 

\item The spectra of the regions `E', `F' and `G'  show molecular absorption bands at at 1.17-1.20\,$\mu$m, possibly originated from evolved stars at the TP-AGB phase.

\item The near-IR emission lines of \feii\ and \hm\ from the star forming regions  have as main ionizing source young stars with 5--6 Myr of age. 

\item The gas velocities at locations beyond the inner 0\farcs45 ($\sim$\,220 pc) are dominated by rotation, being well reproduced by a disk model assuming circular orbits in the plane of the galaxy. At the nucleus, the gas kinematics suggests  two components: one due to the disk of the galaxy, with $V_{\rm LOS}$ similar to the systemic velocity of the galaxy (7391$\pm$11\,km\,s$^{-1}$) and another blueshifted by 300-500~\kms, attributed to a nuclear outflow.

\item The EqW values of the \br\ emission line suggest two age sequences, one from region `A' to `D' and another from region `F' to `G'. These age sequences favor the `pearls on a string' scenario for the origin of the CNSFRs in Mrk~42, although the presence of signatures of TP-AGB stars in some of the regions undermines this interpretation. 

\item The FWHM of the broad component of \pb\ emission line is $\sim$\,1480\,\kms, being larger than the values previously observed for H$\beta$. Using this value we obtain a mass of $M_{BH} = 2.5^{+0.7}_{-0.6}\times10^{6}$~M$_{\odot}$ for the central supermassive black hole.

\end{itemize}

\section*{Acknowledgements}
We thank the anonymous referee for valuable suggestions which helped to improve the paper.
This work is based on observations obtained at the Gemini Observatory, 
which is operated by the Association of Universities for Research in Astronomy, Inc., under a cooperative agreement with the 
NSF on behalf of the Gemini partnership: the National Science Foundation (United States), the Science and Technology 
Facilities Council (United Kingdom), the National Research Council (Canada), CONICYT (Chile), the Australian Research 
Council (Australia), Minist\'erio da Ci\^encia e Tecnologia (Brazil) and south-eastCYT (Argentina). 
This research has made use of the NASA/IPAC Extragalactic Database (NED) which is operated by the Jet Propulsion Laboratory, California Institute of Technology, under contract with the National Aeronautics and Space Administration.
We acknowledge the usage of the HyperLeda database (http://leda.univ-lyon1.fr).
The authors acknowledge support from CNPq and  FAPERGS. 
O.L.D. is grateful to FAPESP (2016/04728-7) and CNPQ (306744/2014-7).

{}


\begin{thebibliography}{}



\bibitem[\protect\citeauthoryear{Albareti et al.}{2017}]{albareti17} Albareti, F. D et al., 2017, ApJS, 233, 25.

\bibitem[\protect\citeauthoryear{Baldwin, Phillips \& Terlevich}{1981}]{bpt81} Baldwin, J. A., Phillips, M. M., Terlevich, R., 1981, PASP, 93, 5

\bibitem[\protect\citeauthoryear{Bertola et al.}{1991}]{bertola} Bertola, F., Bettoni, D., Danziger, J., et al. 1991, ApJ, 373, 369. 

\bibitem[\protect\citeauthoryear{Bian \& Zhao}{2003}]{bian03} Bian W., Zhao Y., 2003, MNRAS, 343, 164.

\bibitem[\protect\citeauthoryear{Black \& van Dishoeck}{1987}]{black87} Black, J. H., \& van Dishoeck, E. F.  1987, ApJ, 322, 412.

\bibitem[\protect\citeauthoryear{Blanton et al.}{2017}]{blanton17} Blanton, M. R. et al., 2017, AJ, 154, 28.

\bibitem[\protect\citeauthoryear{B{\"o}ker et al.}{2008}]{Boker2008} B{\"o}ker, T., Falc{\'o}n-Barroso, J., Schinnerer, E., Knapen, J. H., \& Ryder, S. 2008, AJ, 135, 479. 

\bibitem[\protect\citeauthoryear{Brum et al.}{2017}]{brum17} Brum, C., Riffel, R. A., Storchi-Bergmann, T., Robinson, A., Schnorr-Muller, A., Lena, D., 2017, MNRAS, accepted.

\bibitem[\protect\citeauthoryear{Cardelli, Clayton \& Mathis}{1989}]{cardelli89} Cardelli,J. A., Clayton, G. C. \& Mathis, J. S., 1989, ApJ, 345,245.



\bibitem[\protect\citeauthoryear{Colina \& Arribas}{1999}]{colina99} Colina, L., Arribas, S., 1999, ApJ, 514, 637.


\bibitem[\protect\citeauthoryear{Colina et al.}{2002}]{colina02} Colina, L., Gonzalez Delgado, R., Mas-Hesse, J.~M., \& Leitherer, C.\ 2002, ApJ, 579, 545. 

\bibitem[\protect\citeauthoryear{Colina et al.}{2015}]{colina15} Colina L., Piqueras-Lopez J., Arribas S., R. Riffel, Rodriguez-Ardila, Pastoriza, M. G., Storchi-Bergmann T., Alonso-Herrero \& Sales D., 2015, A\&A, 578, 48.

\bibitem[\protect\citeauthoryear{Copetti et al.}{1986}]{copetti1986} Copetti, M.~V.~F., Pastoriza, M.~G., \& Dottori, H.~A. 1986, A\&A, 156, 111. 

\bibitem[\protect\citeauthoryear{Davies et al.}{2005}]{davies05} Davies, R. I., I., Sternberg, A., Lehnert, M. D., \&  Tacconi-Garman, L. E., 2005, ApJ, 633, 105. 

\bibitem[\protect\citeauthoryear{Davies et al.}{2007}]{davies07} Davies, R. I., M{\"u}ller S{\'a}nchez, F., Genzel, R., et al., 2007, ApJ, 671, 1388. 

\bibitem[\protect\citeauthoryear{Deo, Crenshaw \& Kraemer}{2006}]{deo06} Deo, R. P.; Crenshaw, D. M.; Kraemer, S. B., 2006, AJ, 132, 321.

\bibitem[\protect\citeauthoryear{D{\'{\i}}az et al.}{2007}]{diaz07} D{\'{\i}}az, {\'A}.~I., Terlevich, E., Castellanos, M., \& H{\"a}gele, G.~F.\ 2007, MNRAS, 382, 251. 

\bibitem[\protect\citeauthoryear{Diniz et al.}{2015}]{n2110} Diniz, M. R., Riffel, R. A., Storchi-Bergmann, T., Winge, C., 2015, MNRAS, 453, 1727.

\bibitem[\protect\citeauthoryear{Dottori}{1981}]{dottori1981} Dottori, H.~A. 1981, Ap\&SS, 80, 267. 

\bibitem[\protect\citeauthoryear{Dopita \& Sutherland}{1996}]{Dopita1996} Dopita, M. A., \& Sutherland, R. S., 1996, ApJS, 102, 161 

\bibitem[\protect\citeauthoryear{Dors et al.}{2008}]{dors08} Dors, O. L., Jr., Storchi-Bergmann, T., Riffel, R. A., \& Schimdt, A. A., 2008, A\&A,, 482, 59. 

\bibitem[\protect\citeauthoryear{Dors et al.}{2012}]{dors12} Dors, O. L., Jr., Riffel, R. A., Cardaci, M. V., et al., 2012, MNRAS, 422, 252. 


\bibitem[\protect\citeauthoryear{Elmegreen}{1994}]{Elmegreen1994} Elmegreen, B. G. 1994, ApJL, 425, L73.

\bibitem[\protect\citeauthoryear{Falco et al.}{1999}]{falco1999} Falco, E.~E., Kurtz, M.~J., Geller, M.~J., et al., 1999, PASP, 111, 438. 

\bibitem[\protect\citeauthoryear{Falc{\'o}n-Barroso et al.}{2014}]{falcon2014} Falc{\'o}n-Barroso, J., Ramos Almeida, C., B{\"o}ker, T., et al. 2014, MNRAS, 438, 329. 

\bibitem[\protect\citeauthoryear{Fathi et al.}{2006}]{fathi06} Fathi, K., Storchi-Bergmann, T., Riffel, R. A., Winge, C., Axon, D. J.,  Robinson, A., Capetti, A., \& Marconi, A., 2006, ApJl, 641, L25.

\bibitem[\protect\citeauthoryear{Ferrarese \& Merritt}{2000}]{Ferrarese2000} Ferrarese, L., \& Merritt, D.\ 2000, ApJL, 539, L9. 

\bibitem[\protect\citeauthoryear{Forbes \& Ward}{1993}]{forbes93} Forbes, D. A. \& Ward, M. J.  1993, ApJ, 416, 150.

\bibitem[\protect\citeauthoryear{Galliano \& Alloin}{2008}]{galliano08} Galliano, E., \& Alloin, D. 2008, A\&A, 487, 519.

\bibitem[\protect\citeauthoryear{Gebhardt et al.}{2000}]{Gebhardt2000} Gebhardt, K., Bender, R., Bower, G., et al. 2000, ApJL, 539, L13. 


\bibitem[\protect\citeauthoryear{Goodrich}{1989}]{goodrich89} Goodrich, R.W. 1989, ApJ 342, 224


\bibitem[\protect\citeauthoryear{Hollenbach \& McKee}{1989}]{hollenbach89} Hollenbach, D., \& McKee, C. F., 1989, ApJ, 342, 306.

\bibitem[\protect\citeauthoryear{Hunt \& Malkan}{1999}]{hunt1999} Hunt, L.~K., \& Malkan, M.~A., 1999, ApJ, 516, 660. 

\bibitem[\protect\citeauthoryear{Hunt et al.}{1999}]{hunt99b} Hunt, L. K., Malkan, M. A., Rush, B., Bicay, M. D., Nelson, B. O., Stanga, R. M., Webb, W., 1999, ApJS, 125, 349.


\bibitem[\protect\citeauthoryear{Jogee et al.}{2005}]{Jogee2005} Jogee, S., Scoville, N., \& Kenney, J. D. P. 2005, ApJ, 630, 837. 


\bibitem[\protect\citeauthoryear{Kennicutt}{1989}]{kennicutt89}  Kennicutt, R. C., Keel, W., \& Blaha, C. A. 1989, ApJ, 97, 1022


\bibitem[\protect\citeauthoryear{Kennicutt}{1998}]{kennicutt98} Kennicutt, R. C., Jr., 1998, ARA\&A, 36, 189. 

\bibitem[\protect\citeauthoryear{Kim et al.}{2010}]{kim10}  Kim, D., Im, M., \& Kim, M. 2010, ApJ, 724, 386


\bibitem[\protect\citeauthoryear{Kim et al.}{2015}]{kim2015} Kim, D., Im, M., Glikman, E., Woo, J.-H., \& Urrutia, T. 2015, ApJ, 812, 66. 



\bibitem[\protect\citeauthoryear{Knapen}{2005}]{Knapen2005} Knapen, J. H., 2005, A\&A, 429, 141. 

\bibitem[\protect\citeauthoryear{Kotilainen et al.}{2000}]{Kotilainen} Kotilainen, J.~K., Reunanen, J., Laine, S., \& Ryder, S.~D. 2000, A\&A, 353, 834. 


\bibitem[\protect\citeauthoryear{Lan{\c c}on \& Mouhcine}{2002}]{lanco2002} Lan{\c c}on, A., \& Mouhcine, M., 2002, A\&A, 393, 167. 

\bibitem[\protect\citeauthoryear{Lamperti et al.}{2017}]{lamperti17} Lamperti, I. et al., 2017, MNRAS, 467, 540. 

\bibitem[\protect\citeauthoryear{Larkin et al.}{1998}]{larkin98} Larkin, J. E., Armus, L., Knop, R. A., Soifer, B. T., Matthews, K., 1998, ApJS, 114, 59.	

\bibitem[\protect\citeauthoryear{Makarov et al.}{2014}]{makarov14} Makarov, D., Prugniel, P., Terekhova, N., Courtois, H., Vauglin, I., 2014, A\&A, 570, 13

\bibitem[\protect\citeauthoryear{Malkan et al.}{1998}]{malkan98} Malkan, M. A., Gorjian, V., \& Tam, R. 1998, ApJS, 117, 25. 

\bibitem[\protect\citeauthoryear{Maloney, Hollenbach \& Tielens}{1996}]{maloney96} Maloney, P. R.,  Hollenbach, D. J.,  \& Tielens, A. G. G. M., 1996, ApJ, 466, 561.

\bibitem[\protect\citeauthoryear{May et al.}{2016}]{may16}  May, D., Steiner, J. E., Ricci, T. V., Menezes, R. B., Andrade, I. S., 2016, MNRAS, 457, 949.


\bibitem[\protect\citeauthoryear{Maraston}{2005}]{maraston05} Maraston, C. 2005, MNRAS, 362, 799. 

\bibitem[\protect\citeauthoryear{Markwardt et al.}{2009}]{mark09} Markwardt C. B., 2009, in Bohlender D. A., Durand D., Dowler P., eds, ASP Conf. Ser. Vol. 411, Astronomical Data Analysis Software and Systems XVIII. Astron. Soc. Pac., San Francisco, p. 251

\bibitem[\protect\citeauthoryear{Martins et al.}{2013a}]{martins13} Martins, L.~P., Rodr{\'{\i}}guez-Ardila, A., Diniz, S., Gruenwald, R., \& de Souza, R. 2013, MNRAS, 431, 1823.

\bibitem[\protect\citeauthoryear{Martins et al.}{2013b}]{martins13b} Martins, L.~P., Rodr{\'{\i}}guez-Ardila, A., Diniz, S., Riffel, R., \& de Souza, R. 2013, MNRAS, 435, 2861.

\bibitem[\protect\citeauthoryear{Mazzuca et al.}{2008}]{mazzuca2008} Mazzuca, L.~M., Knapen, J.~H., Veilleux, S., \& Regan, M.~W.\ 2008, ApJS, 174, 337-365. 

\bibitem[\protect\citeauthoryear{McGregor et al.}{2003}]{mcgregor03} McGregor, P. J. et al., 2003, Proceedings of the SPIE, 4841, 1581.



\bibitem[\protect\citeauthoryear{Morgan}{1958}]{morgan1958} Morgan, W.~W. 1958, PASP, 70, 364. 

\bibitem[\protect\citeauthoryear{M{\"u}ller S\'anchez et al.}{2009}]{ms09}  M\"uller S\'anchez, F., Davies, R. I., Genzel, R., Tacconi, L. J., Eisenhauer, F., Hicks, E. K. S., Friedrich, S., Sternberg, A., 2009, ApJ, 691, 749.

\bibitem[\protect\citeauthoryear{Mu{\~n}oz Mar{\'{\i}}n et al.}{2007}]{Munoz2007} Mu{\~n}oz Mar{\'{\i}}n, V. M., Gonz{\'a}lez Delgado, R. M., Schmitt, H. R., et al. 2007, AJ, 134, 648. 

\bibitem[\protect\citeauthoryear{Nikolajuk, Czerny \& Gurynowicz}{2009}]{nikolajuk09} Nikolajuk, M., Czerny, B., Gurynowicz, P., 2009, MNRAS, 394, 2141.

\bibitem[\protect\citeauthoryear{Norman \& Scoville}{1988}]{norman88} Norman, C., \& Scoville, N.\ 1988, ApJ, 332, 124.

\bibitem[\protect\citeauthoryear{Osterbrock \& Ferland}{2006}]{osterbrock06} Osterbrock, D. E. \& Ferland, G. J., 2006, Astrophysics of Gaseous Nebulae and Active Galactic Nuclei, Second Edition, University Science Books, Mill Valley, California.

\bibitem[\protect\citeauthoryear{Perry \& Dyson}{1985}]{perry85} Perry, J. J., \& Dyson, J. E., 1985, MNRAS, 213, 665.

\bibitem[\protect\citeauthoryear{Ramos Almeida,   P\'erez Garc\'\i a \& Acosta-Pulido}{2009}]{ra09} Ramos Almeida, C., P\'erez Garc\'\i a, A. M.,
 \& Acosta-Pulido, J. A., 2009, ApJ, 694, 1379.

\bibitem[\protect\citeauthoryear{Rayner et al.}{2009}]{rayner09} Rayner, J. T., Cushing, M. C., \& Vacca, W. D., 2009, ApJS, 185, 289.

\bibitem[\protect\citeauthoryear{Reunanen, Kotilainen \& Prieto}{2002}]{reunanen02} Reunanen, J., Kotilainen, J. K., \& Prieto, M. A., 2002, MNRAS, 331, 154. 

\bibitem[\protect\citeauthoryear{Ricci, Steiner \& Menezes}{2014}]{ricci14} Ricci, T. V.; Steiner, J. E.; Menezes, R. B., MNRAS, 440, 2419.

\bibitem[\protect\citeauthoryear{Riffel et al.}{2006}]{eso428} Riffel, Rogemar A., Sorchi-Bergmann, T., Winge, C., Barbosa, F. K. B., 2006, MNRAS, 373, 2.

\bibitem[\protect\citeauthoryear{Riffel et al.}{2008}]{n4051} Riffel, Rogemar A., Storchi-Bergmann, T., Winge, C., McGregor, P. J., Beck, T., Schmitt, H. 2008, MNRAS, 385, 1129.

\bibitem[\protect\citeauthoryear{Riffel et al.}{2009}]{n7582} Riffel, Rogemar A., Storchi-Bergmann, T., Dors, O. L., Winge, C., 2009, MNRAS, 393, 783.

\bibitem[\protect\citeauthoryear{Riffel}{2010}]{profit} Riffel, Rogemar A., 2010, Ap\&SS, 327, 239.


\bibitem[\protect\citeauthoryear{Riffel \& Storchi-Bergmann}{2011a}]{mrk1066} Riffel, Rogemar A. \& Storchi-Bergmann, T., 2011, MNRAS, 411, 469.


\bibitem[\protect\citeauthoryear{Riffel, Storchi-Bergmann \& Winge}{2013}]{mrk79} Riffel, R. A., Storchi-Bergmann, T., Winge, C., 2013, 430, 2249.

\bibitem[\protect\citeauthoryear{Riffel et al.}{2010}]{mrk1066a} Riffel, Rogemar A., \& Storchi-Bergmann, T. \& Nagar, N. M., 2010, MNRAS, 404, 166.

\bibitem[\protect\citeauthoryear{Riffel et al.}{2016}]{Riffel2016} Riffel, Rogemar A., Colina, L., Storchi-Bergmann, T., Piqueras L\'opez J., Arribas, S., Riffel, R., Pastoriza, M., Sales, Dinalva A., Dametto, N. Z., Labiano, A. \& Davies, R. I., 2016, MNRAS, 461, 4192.

\bibitem[\protect\citeauthoryear{Riffel et al.}{2018}]{riffel18} Riffel, R. A. et al., 2018, MNRAS, 474, 1373.


\bibitem[\protect\citeauthoryear{Riffel et al.}{2007}]{rogerio07} Riffel, R., Pastoriza, M. G., Rodriguez-Ardila, A., Maraston, C., 2007, ApJ, 659, 103.


\bibitem[\protect\citeauthoryear{Riffel et al.}{2011}]{rogerio2011} Riffel, R., Bonatto, C., Cid Fernandes, R., Pastoriza, M.~G., \& Balbinot, E. 2011, MNRAS, 411, 1897. 

\bibitem[\protect\citeauthoryear{Riffel et al.}{2013}]{rogerio2013} Riffel, R., Rodr{\'{\i}}guez-Ardila, A., Aleman, I., et al. 2013, MNRAS, 430, 2002. 

\bibitem[\protect\citeauthoryear{Riffel et al.}{2015}]{rogerio15} Riffel, R. et al., 2015, MNRAS, 450, 3069.

\bibitem[\protect\citeauthoryear{Rodr\'\i guez-Ardila et al.}{2004}]{ardila04} Rodr\'\i guez-Ardila, A.,  Pastoriza, M. G., Viegas, S., Sigut, T. A. A., \& Pradhan, A. K., 2004,  A\&A, 425, 457.

\bibitem[\protect\citeauthoryear{Rodr\'\i guez-Ardila, Riffel \& Pastoriza}{2005}]{ardila05} Rodr\'\i guez-Ardila, A., Riffel, R., \& Pastoriza, M. G. 2005,  MNRAS, 364, 1041.

\bibitem[\protect\citeauthoryear{Sakamoto et al.}{1999}]{Sakamoto1999} Sakamoto, K., Okumura, S. K., Ishizuki, S., \& Scoville, N. Z. 1999, ApJ, 525, 691. 

\bibitem[\protect\citeauthoryear{Sani al.}{2010}]{sani10} Sani, E., Lutz, D., Risaliti, G., Netzer, H., Gallo, L. C., Trakhtenbrot, B., Sturm, E., Boller, T., 2010, MNRAS, 403, 1246.


\bibitem[\protect\citeauthoryear{Sch{\"o}nell et al.}{2017}]{n5548} Sch{\"o}nell, A.~J., Riffel, R.~A., Storchi-Bergmann, T., \& Riffel, R. 2017, MNRAS, 464, 1771. 

\bibitem[\protect\citeauthoryear{Schnorr-M\"uller al.}{2014a}]{sm14a} Schnorr-M\"uller, Allan; Storchi-Bergmann, T.,  Nagar, N. M.,  Lena, D., Riffel, R. A., Couto, G. S., 2014b, 2014a, MNRAS, 438, 3322.

\bibitem[\protect\citeauthoryear{Schnorr-M\"uller al.}{2014b}]{sm14b} Schnorr-M\"uller, Allan; Storchi-Bergmann, T.,  Nagar, N. M., Robinson, A., Ferrari, F.,  MNRAS, 437, 1708.

\bibitem[\protect\citeauthoryear{Schnorr--M{\"u}ller et al.}{2017}]{sm17} Schnorr M{\"u}ller A., Storchi-Bergmann T., Ferrari, F.,  Nagar, N. M., 2017, MNRAS, 466, 4370.

\bibitem[\protect\citeauthoryear{Schwarz}{1981}]{schwarz81} Schwarz, M. P. 1981, ApJ, 247, 77

\bibitem[\protect\citeauthoryear{Scoville et al.}{1982}]{scoville1982} Scoville, N. Z.,  Hall, D. N. B., Ridgway, S. T., \& Keinmann, S. G. 1982, ApJ, 253, 136.


\bibitem[\protect\citeauthoryear{S{\'e}rsic \& Pastoriza}{1967}]{sersic1967} S{\'e}rsic, J.~L., \& Pastoriza, M. 1967, PASP, 79, 152. 

\bibitem[\protect\citeauthoryear{Shi, Gu \& Peng}{2006}]{shi06} Shi, L., Gu, Q. S., \& Peng, Z. X. 2006, A\&A, 450, 15.

\bibitem[\protect\citeauthoryear{Sim\~oes et al.}{2007}]{simoes2007} Sim\~oes Lopes R. D., Storchi-Bergmann T., de F\'atima Saraiva M. \& Martini P., 2007, ApJ, 655, 718.

\bibitem[\protect\citeauthoryear{Simpson et al.}{1996}]{simpson96} Simpson, C., Forbes, D. A., Baker, A. C., \& Ward, M. J. 1996, MNRAS, 283, 777.


\bibitem[\protect\citeauthoryear{Storchi-Bergmann et al.}{2007}]{sb07} Storchi-Bergmann, T., Dors Jr., O.,  Riffel,  R. A., Fathi, K.,  Axon, D. J., \& Robinson, A., 2007, ApJ, x,  x.

\bibitem[\protect\citeauthoryear{Storchi-Bergmann}{2008}]{sb08} Storchi-Bergmann, T. 2008, Revista Mexicana de Astronomia y Astrofisica Conference Series, 32, 139. 



\bibitem[\protect\citeauthoryear{van de Ven \& Fathi}{2010}]{vandeVen10} van de Ven, G., \& Fathi, K., 2010, ApJ, 723, 767.

\bibitem[\protect\citeauthoryear{van der Laan et al.}{2015}]{van-der-lan2015} van der Laan, T. P. R., Armus, L., Beirao, P., et al. 2015, A\&A, 575, A83. 

\bibitem[\protect\citeauthoryear{Wang \& Lu}{2001}]{wang01} Wang T., Lu Y., 2001, A\&A, 377, 52.

\bibitem[\protect\citeauthoryear{Wilson et al.}{1991}]{Wilson} Wilson, A.~S., Helfer, T.~T., Haniff, C.~A., \& Ward, M.~J., 1991, ApJ, 381, 79. 

\bibitem[\protect\citeauthoryear{Wold \& Galliano}{2006}]{wold06} Wold, M., Galliano, E., 2006, MNRAS, 369, 47.

\bibitem[\protect\citeauthoryear{Woo et al.}{2015}]{woo15} Woo, J.-H., Yoon, Y., Park, S., Park, D., \& Kim, S. C. 2015, ApJ, 801, 38





  














\end{thebibliography}
\end{document}